\documentclass[runningheads]{llncs}

\usepackage{tikz}
\usepackage{amsmath}
\usepackage{listings}
\usepackage{xspace}
\usepackage[colorlinks,citecolor=.]{hyperref}
\usepackage{bigstrut}

\usepackage{booktabs}

\usepackage{subfigure}
\usepackage{url}
\usepackage{mathtools}
\usepackage{amsfonts}
\usepackage{amssymb}
\usepackage{multirow}
\usepackage[ruled,vlined]{algorithm2e}
\usepackage{algpseudocode}
\usepackage{graphicx}
\usepackage{enumitem}
\usepackage{numprint}
\usepackage[normalem]{ulem}
\usepackage{tcolorbox}
\usepackage{xurl}
\usepackage{xstring}

\usepackage{color}
\definecolor{lightgray}{rgb}{.9,.9,.9}
\definecolor{darkgray}{rgb}{.4,.4,.4}
\definecolor{purple}{rgb}{0.65, 0.12, 0.82}

\lstdefinelanguage{JavaScript}{
  keywords={typeof, new, true, false, catch, function, return, null, catch, switch, var, if, in, while, do, else, case, break},
  keywordstyle=\color{blue}\bfseries,
  ndkeywords={class, export, boolean, throw, implements, import, this},
  ndkeywordstyle=\color{darkgray}\bfseries,
  identifierstyle=\color{black},
  sensitive=false,
  comment=[l]{//},
  morecomment=[s]{/*}{*/},
  commentstyle=\color{purple}\ttfamily,
  stringstyle=\color{red}\ttfamily,
  morestring=[b]',
  morestring=[b]"
}

\newcommand{\etal}{\textit{et al.\ }}
\npthousandsep{,}

\usepackage{tikz}
\newcommand*\circled[1]{\tikz[baseline=(char.base)]{
            \node[shape=circle,draw,inner sep=0.5pt] (char) {#1};}}

\usepackage{lmodern}

\newcommand{\Loan}{{\color{blue}\normalfont\texttt{Loan}}}
\newcommand{\Repay}{{\color{blue}\normalfont\texttt{Repay}}}
\newcommand{\SwapXY}{{\color{blue}\normalfont\texttt{Swap$\mathsf{X}$for$\mathsf{Y}$}}}
\newcommand{\SwapYX}{{\color{blue}\normalfont\texttt{Swap$\mathsf{Y}$for$\mathsf{X}$}}}
\newcommand{\CollateralizedBorrow}{{\color{blue}\normalfont\texttt{CollateralizedBorrow}}}
\newcommand{\CollateralizedRepay}{{\color{blue}\normalfont\texttt{CollateralizedRepay}}}
\newcommand{\MarginShort}{{\color{blue}\normalfont\texttt{MarginShort}}}
\newcommand{\ConvertXY}{{\color{blue}\normalfont\texttt{Convert$\mathsf{X}$to$\mathsf{Y}$}}}
\newcommand{\SellXY}{{\color{blue}\normalfont\texttt{Sell$\mathsf{X}$for$\mathsf{Y}$}}}
\newcommand{\point}[1]{\par\noindent\textbf{#1:}\xspace}

\newcommand{\ImprovementPumpAndArbitrage}{829.5k~USD\xspace}
\newcommand{\RatioPumpAndArbitrage}{2.37$\times$\xspace}

\newcommand{\ImprovementOracleManipulation}{1.1M~USD\xspace}
\newcommand{\RatioOracleManipulation}{1.73$\times$\xspace}

\newcommand{\states}{\par\noindent{\sf State:} }
\newcommand{\transitions}{\par\noindent{\sf Transitions:} }

\newcommand{\block}[1]{\href{https://etherscan.io/block/#1}{#1}\xspace}
\newcommand{\transaction}[1]{\href{https://etherscan.io/tx/#1}{\wrapletters{#1}}\xspace}

\newcommand{\address}[1]{\href{https://etherscan.io/address/#1}{\wrapletters{#1}}\xspace}

\newcommand*\wrapletters[1]{\wr@pletters#1\@nil}
\def\wr@pletters#1#2\@nil{#1\allowbreak\if&#2&\else\wr@pletters#2\@nil\fi}

\usepackage{pdflscape}
\usepackage{rotating}
\usepackage{array}
\newcolumntype{L}{>{\centering\arraybackslash}m{3cm}}

\usepackage{caption}
\usepackage{wrapfig}

\begin{document}

\title{Attacking the DeFi Ecosystem with Flash Loans for Fun and Profit}


\author{Kaihua Qin \and
Liyi Zhou \and
Benjamin Livshits \and Arthur Gervais}
\authorrunning{K. Qin et al.}
%
\institute{Imperial College London, United Kingdom\\
\email{\{kaihua.qin,liyi.zhou,b.livshits,a.gervais\}@imperial.ac.uk}}

\maketitle


\begin{abstract}
Credit allows a lender to loan out surplus capital to a borrower. In the traditional economy, credit bears the risk that the borrower may default on its debt, the lender hence requires upfront collateral from the borrower, plus interest fee payments. Due to the atomicity of blockchain transactions, lenders can offer \emph{flash loans}, i.e., loans that are only valid within one transaction and must be repaid by the end of that transaction. This concept has lead to a number of interesting attack possibilities, some of which were exploited in February~2020.

This paper is the first to explore the implication of transaction atomicity and flash loans for the nascent decentralized finance~(DeFi) ecosystem. We show quantitatively how transaction atomicity increases the arbitrage revenue. We moreover analyze two existing attacks with ROIs beyond 500k\%. We formulate finding the attack parameters as an \emph{optimization problem} over the state of the underlying Ethereum blockchain and the state of the DeFi ecosystem. We show how malicious adversaries can efficiently maximize an attack profit and hence damage the DeFi ecosystem further. Specifically, we present how two previously executed attacks can be ``boosted'' to result in a profit of ~\ImprovementPumpAndArbitrage and~\ImprovementOracleManipulation, respectively, which is a boost of~\RatioPumpAndArbitrage and~\RatioOracleManipulation, respectively.
\end{abstract}

\setlength{\textfloatsep}{0.5\baselineskip plus 0.2\baselineskip minus 0.4\baselineskip}

\section{Introduction}
\label{sec:intro}
A central component of our economy is \emph{credit}: to foster economic growth, market participants can borrow and lend assets to each other. If credit creates new and sustainable value, it can be perceived as a positive force. Abuse of credit, however, necessarily entails negative future consequences. Excessive debt can lead to a debt default --- i.e., a borrower is no longer capable to repay the loan plus interest payment. This leads us to the following intriguing question:
What if it were possible to offer credit without bearing the risk that the borrower does not pay back the debt?
Such a concept appears impractical in the traditional financial world. No matter how small the borrowed amount, and how short the loan term, the risk of the borrower defaulting remains. If one were absolutely certain that a debt would be repaid, one could offer loans of massive volume~--~or lend to individuals independently of demographics and geographic location, effectively providing capital to rich and poor alike.

Given the peculiarities of blockchain-based smart contracts, \emph{flash loans} emerged. Blockchain-based smart contracts allow to programmatically enforce the atomic execution of a transaction. A flash loan is a loan that is only valid within one atomic blockchain transaction. Flash loans fail if the borrower does not repay its debt before the end of the transaction borrowing the loan. That is because a blockchain transaction can be reverted during its execution if the condition of repayment is not satisfied. Flash loans yield three novel properties, absent in traditional finance:
\begin{itemize}
    \item\textbf{No debt default risk:} A lender offering a flash loan bears no risk that the borrower defaults on its debt\footnote{Besides the risk of smart contract vulnerabilities.}. Because a transaction and its instructions must be executed atomically, a flash loan is not granted if the transaction fails due to a debt default.
    \item\textbf{No need for collateral:} Because the lender is guaranteed to be paid back, the lender can issue credit without upfront collateral from the borrower: a flash loan is non-collateralized.
    \item\textbf{Loan size:} Flash loans can be taken from public smart contract-governed liquidity pools. Any borrower can borrow the entire pool at any point in time. As of September~2020, the largest flash loan pool Aave~\cite{aave} offers in excess of~$1$B~USD~\cite{Aavewatc13:online}.
\end{itemize}

To the best of our knowledge, this is the first paper that investigates flash loans. \textbf{This paper makes the following contributions:}

\begin{itemize}
    \item \textbf{Flash loan usage analysis.}
    We provide a comprehensive overview of how and where the technique of flash loans can and is utilized. At the time of writing, flash loan pool sizes have reached more than $1$B~USD.
    
    \item \textbf{Post mortem of existing attacks.}
    We meticulously dissect two events where talented traders realized a profit of each about~$350$k USD and~$600$k USD with two independent flash loans: a \emph{pump and arbitrage} from the~15th of February~2020 and an \emph{oracle manipulation} from the~18th of February~2020.
    
    \item \textbf{Attack parameter optimization framework.}
    Given the interplay of six DeFi systems, covering exchanges, credit/lending, and margin trading, we provide a framework to quantify the parameters that yield the maximum revenue an adversary can achieve, given a specific trading attack strategy.
    We show that an adversary can maximize the attack profit efficiently (in less than $13$ms) due to the atomic transaction property.

    \item \textbf{Quantifying opportunity loss.}
    We show how the presented flash loan attackers have forgone the opportunity to realize a profit exceeding~\ImprovementPumpAndArbitrage and~\ImprovementOracleManipulation, respectively. We realize this by finding the optimal adversarial parameters the trader should have employed, using a parametrized optimizer. We experimentally validate the opportunity loss on a locally deployed blockchain mirroring the attacks' respective blockchain state.
    
    \item \textbf{Impact of transaction atomicity on arbitrage.}
    We show quantitatively how atomicity reduces the risk of revenue from arbitrage. Specifically, by analyzing 6.4M transactions, we find that the expected arbitrage reward decreases by~$123.49 \pm 1375.32$ USD and~$1.77 \pm 10.59$ USD for the DAI/ETH and MKR/ETH markets respectively when the number of intermediary transactions reaches~$5,000$.
\end{itemize}

\point{Paper organization}
The remainder of the paper is organized as follows. Section~\ref{sec:background} elaborates on the DeFi background. Section~\ref{sec:postmortem} dissects two known flash loan attacks. Section~\ref{sec:framework} proposes a framework to optimize the attack revenues and Section~\ref{sec:evaluation} evaluates the framework on the two analyzed attacks. Section~\ref{sec:atomicity} analyses the implications of the atomic transaction property. Section~\ref{sec:discussion} provides a discussion. We conclude the paper in Section~\ref{sec:conclusion}.

\section{Background}
\label{sec:background}
Decentralized ledgers, such as Bitcoin~\cite{bitcoin}, enable the performance of transactions among a peer-to-peer network. At its core, a blockchain is a chain of blocks~\cite{bonneau2015sok,bitcoin}, extended by miners crafting new blocks that contain transactions. Smart contracts~\cite{wood2014ethereum} allow the execution of complicated transactions, which forms the foundation of decentralized finance, a conglomerate of financial cryptocurrency-related protocols. These protocols for instance allow to lend and borrow assets~\cite{makerdao,compoundfinance}, exchange~\cite{dydx,UniswapH70:online}, margin trade~\cite{dydx,bZxAProt64:online}, short and long~\cite{bZxAProt64:online}, and allow to create derivative assets~\cite{compoundfinance}. At the time of writing, the DeFi space accounts for over~$8$B~USD in smart contract locked capital among different providers. The majority of the DeFi platforms operate on the Ethereum blockchain, governed by the Ethereum Virtual Machine (EVM), where the trading rules are governed by the underlying smart contracts. A decentralized exchange is typically referred to as DEX. We refer to the on-chain DeFi actors as traders and distinguish among the two types of traders:
\point{Liquidity Provider} a trader with surplus capital may choose to offer this capital to other traders, e.g., as collateral within a DEX or lending platform.
\point{Liquidity Taker} a trader which is servicing liquidity provider with fees in exchange for accessing the available capital.

\subsection{DeFi Platforms}
We briefly summarize relevant DeFi platforms for this work.



\point{Automated market maker~(AMM) DEX}
While many exchanges follow the limit order book design~\cite{oasis2019,idex2019,kyber}, an alternative exchange design is to collect funds within a liquidity pool, e.g., two pools for an AMM asset pair $X$/$Y$~\cite{UniswapH70:online,kyber}. The state (or depth) of an AMM market $X$/$Y$ is defined as $(x, y)$, where $x$ represents the amount of asset $X$ and $y$ the amount of asset $Y$ in the liquidity pool. Liquidity providers can deposit/withdraw in both assets $X$ and $Y$ to in/decrease liquidity. The simplest AMM mechanism is a constant product market maker, which for an arbitrary asset pair $X$/$Y$, keeps the product $x \times y$ constant during trades. When trading on an AMM exchange, there can be a difference between the expected price and the executed price, termed \emph{slippage}~\cite{Slippage53:online}. Insufficient liquidity and other front-running trades can cause slippage on an AMM~\cite{zhou2020high}.
We assume that a constant product AMM ETH/WBTC market is supplied with~$10$~ETH and $10$~WBTC (i.e., the exchange rate is~$1$~ETH/WBTC). A trader can purchase~$5$~WBTC with~$10$~ETH (cf.\ $10\times10=(10+10)\times(10-5)$) at an effective price of~$2$~ETH/WBTC. Hence, the slippage is $\frac{2-1}{1}=100\%$.

\point{Margin trading}
Trading on margin allows a trader to take under-collateralized loans from the trading platform and trade with these borrowed assets to amplify the profit (i.e., leverage). On-chain margin trading platforms remain in control of the loaned asset (or the exchanged asset) and hence is able to liquidate when the value of the trader's collateral drops too low.

\point{Credit and lending}
With over~$3$B USD total locked value, credit represents one of the most significant recent use-cases for blockchain based DeFi systems. Due to the lack of legal enforcement when borrowers default, they are required to provide between 125\%~\cite{dydx} to 150\%~\cite{makerdao} collateral of an asset $x$ to borrow 100\% of another asset $y$ (i.e., over-collateralization).

\subsection{Reverting EVM State Transitions}
The Ethereum blockchain is in essence a replicated state machine. To achieve a state transition, one applies as input transactions that modify the EVM state following rules encoded within deployed smart contracts. A smart contract can be programmed with the logic of reverting a transaction if a particular condition is not met during execution. The EVM state is \emph{only} altered if a transaction executes successfully, otherwise, the EVM state is reverted to the previous, non-modified state.

\paragraph{Flash Loans.}
Flash loans are possible because the EVM allows the reversion of state changes. A flash loan is only valid within a single transaction and relies on the atomicity of blockchain (and, specifically, EVM) transactions within a single block. 
Flash loans entail two important new financial properties: First, a borrower does not need to provide upfront collateral to request a loan of any size, up to the flash loan liquidity pool amount. Any borrower, willing to pay the required transaction fees (which typically amounts to a few USD) is an eligible borrower. Second, risk-free lending: If a borrower cannot pay back the loan, the flash loan transaction fails. Ignoring smart contract and blockchain vulnerabilities, the lender is hence not exposed to the risks of a debt default.

\subsection{Flash Loan Usage in the Wild}
To our knowledge, the Marble Protocol introduced the concept of flash loans~\cite{marblepr82:online}. Aave~\cite{aave} is one of the first DeFi platforms to widely advertise flash loan capabilities (although others, such as dYdX also allow the non-documented possibility to borrow flash loans) since January 2020. At the time of writing, Aave charges a constant~0.09\% interest fee for flash loans and amassed a total liquidity beyond~$1$B~USD~\cite{Aavewatc13:online}. In comparison, the total volume of U.S. corporation debt reached~$10.5$T~USD in August,~2020~\cite{UScorpor25:online}.

\begin{figure}[htb!]
\begin{center}
\includegraphics[width=0.9\columnwidth]{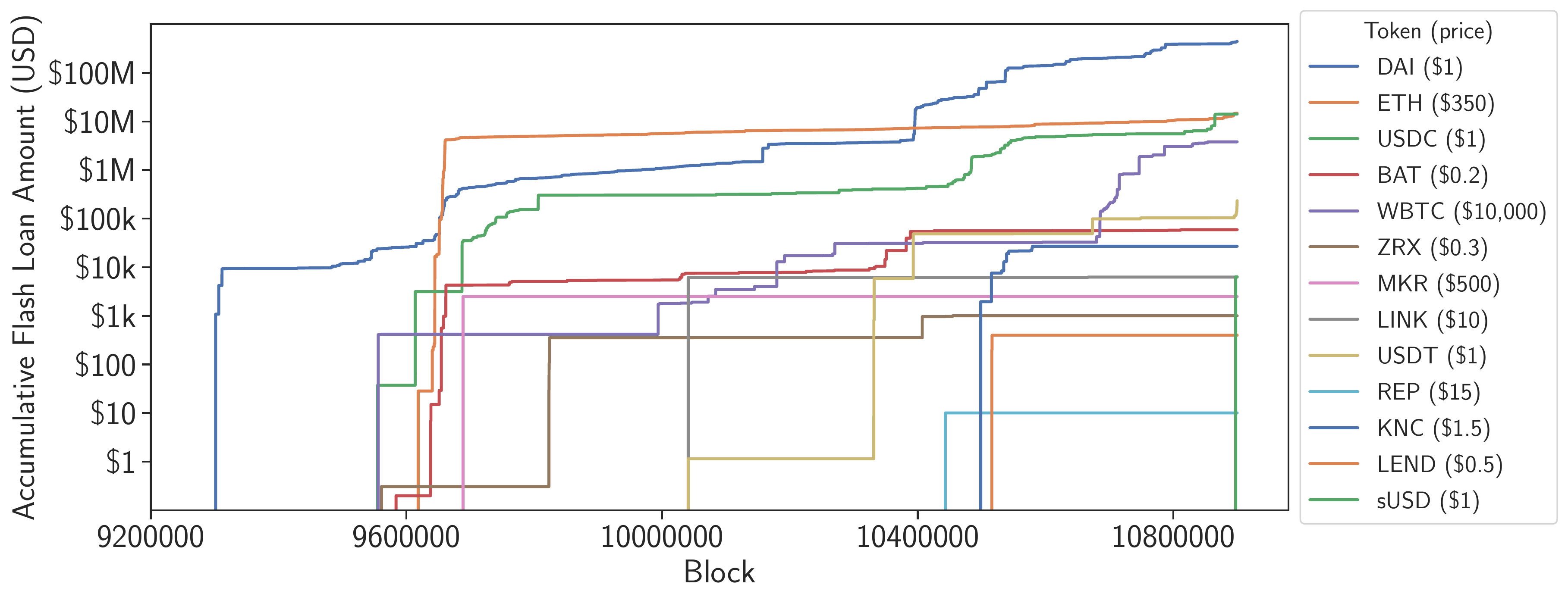}
\end{center}
\vspace{-20pt}
\caption{Accumulative flash loan amounts of $13$ cryptocurrencies on Aave. Note that the y-axis is a logarithmic scale.
}
\label{fig:aave_cum}
\end{figure}

By gathering all blockchain event logs from Aave with a full archive Ethereum node, we find $5,616$ flash loans issued from the Aave smart contract (cf.\ \address{0x398eC7346DcD622eDc5ae82352F02bE94C62d119}) between the~8th of January,~2020 and the~20th of September,~2020. In Figure~\ref{fig:aave_cum}, we show the accumulative flash loan amounts of $13$ different loan currencies. Among them, DAI is the most popular with the accumulative amount of~$447.2$M USD. We inspect and classify the Aave flash loan transactions depending on which platforms the flash loans interact with (cf.~Figure~\ref{fig:usage} in Appendix~\ref{app:classifyingflashloanusecases}). We notice that most flash loans interact with lending/exchange DeFi systems and that the flash loan's transaction costs (i.e., gas) appear significant (at times beyond 4M gas, compared to~21k gas for regular Ether transfer). The dominating use cases are arbitrage and liquidation. Further details are presented in Appendix~\ref{app:classifyingflashloanusecases}.

\begin{figure}[b!]
    \centering
    \includegraphics[width=0.8\columnwidth]{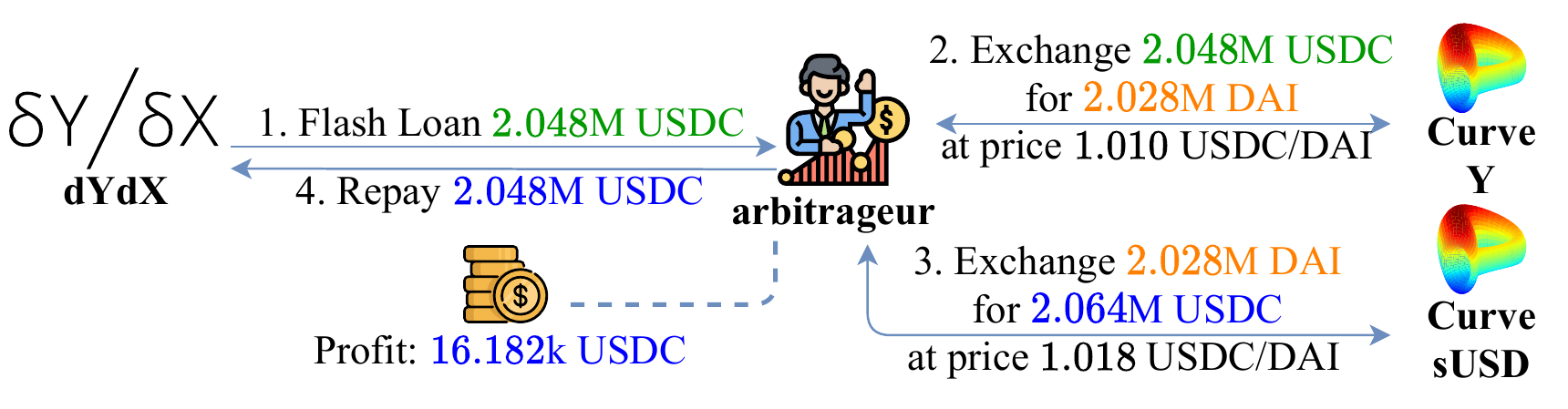}
    \caption{High-level executions of a flash loan based arbitrage transaction \transaction{0xf7498a2546c3d70f49d83a2a5476fd9dcb6518100b2a731294d0d7b9f79f754a}: \emph{(1)} flash loan; \emph{(2)} exchange USDC for DAI in Curve Y pool; \emph{(3)} exchange DAI for USDC in Curve sUSDC pool; \emph{(4)} repay. Note that Curve provides several on-chain cryptocurrency markets, also known as pools.}
    \label{fig:arbitrage-example}
\end{figure}
\point{Flash Loan Arbitrage example}
The value of an asset is typically determined by the demand and supply of the market, across different exchanges. Due to a lack of instantaneous synchronization among exchanges, the same asset can be traded at slightly different prices on different exchanges. \emph{Arbitrage} is the process of exploiting price differences among exchanges for a financial gain~\cite{shleifer1997limits}.
In Figure~\ref{fig:arbitrage-example}, we present, as an example, the execution details of a flash loan based arbitrage transaction on~the 31st of July,~2020. The arbitrageur borrowed a flash loan of~$2.048$M~USDC, performed two exchanges, and realized a profit of~$16.182$k~USDC ($16.182$k~USD). This example highlights how given atomic transactions, a trader can perform arbitrage on different on-chain markets, without the risk that the prices in the DEX would intermediately change. Flash loans moreover remove the currency volatility risk for arbitrageurs. In Section~\ref{sec:atomicity}, we quantify the implications of transaction atomicity on arbitrage risks.

Besides arbitrage, we noticed another two use cases for flash loans: \emph{(i)} wash trading (fraudulent inflation of trading volume), \emph{(ii)} loan collateral swapping (instant swapping from one collateral to another), and also a variation of flash loan, \emph{(iii)} flash minting (the momentarily token in- and decrease of an asset). We elaborate further on these in Appendix~\ref{sec:flash-use-cases} and provide real-world examples.

\subsection{Related work}
There is a growing body of work focusing on various forms of manipulation and financially-driven attacks in cryptocurrency markets.

\point{Crypto Manipulation}
Front-running in cryptocurrencies has been extensively studied~\cite{eskandari2019sok,daian2019flash,breidenbach2018enter,kalodner2015empirical,ConsenSy8:online,zhou2020high}. Remarkably, Daian~\etal~\cite{daian2019flash} introduce the concept of miner extractable value (MEV) and analyze comprehensively the exploitability of ordering blockchain transactions. Our work focuses on flash loans, which qualify as a potential MEV that miners could exploit. Gandal~\etal~\cite{Gandal2018} demonstrate that the unprecedented spike in the USD-{BTC} exchange rate in late 2013 was possibly caused by price manipulation. 
Recent papers focus on the phenomenon of pump-and-dump for manipulating crypto coin prices~\cite{livshits19pump-and-dump,to-the-moon,hamrick2018economics}. 
\point{Smart Contract Vulnerabilities} Several exploits have taken advantage of smart contract vulnerabilities (e.g., the DAO exploit~\cite{Reportof0:online}). The most commonly known smart contract vulnerabilities are re-entrancy, unhandled exceptions, locked ether, transaction order dependency and integer overflow~\cite{Luu2016}. Many tools and techniques, based on fuzzing~\cite{liu2018reguard,jiang2018contractfuzzer,wustholz2019harvey}, static analysis~\cite{tsankov2018securify,brent2018vandal,tikhomirov2018smartcheck}, symbolic execution~\cite{Luu2016,mueller2017mythril,nikolic2018finding}, and formal verification~\cite{bhargavan2016formal,amani2018towards,grishchenko2018semantic,hildenbrandt2017kevm,hirai2017defining}, emerged to detect and prevent these vulnerabilities. In this work, we focus on DeFi economic security, which might not result from a single contract vulnerability and could involve multiple DeFi platforms.

\section{Flash Loan Post-Mortem}
\label{sec:postmortem}
\newcommand{\step}[1]{\circled{\textsf{#1}}\xspace}

Flash loans enable anyone to have instantaneous access to massive capital.
This section outlines how that can have negative effects, as we explain two attacks facilitated by flash loans yielding an ROI beyond~500k\%. We evaluate the proposed DeFi attack optimization framework (cf.\ Section~\ref{sec:framework}) on these two analyzed attacks (cf.\ Section~\ref{sec:evaluation}).

\subsection{Pump Attack and Arbitrage (PA\&A)}
On the~$15$th of February,~$2020$, a flash loan transaction (cf.\ \transaction{0xb5c8bd9430b6cc87a0e2fe110ece6bf527fa4f170a4bc8cd032f768fc5219838} at an ETH price of $264.71$~USD/ETH), followed by $74$ transactions, yielded a profit of $1,193.69$ ETH ($350$k USD) given a transaction fee of~$132.36$ USD (cumulative~$50,237,867$ gas, $0.5$ ETH). We show in Section~\ref{sec:optimizing-existing-attacks1} that the adversarial parameters were not optimal, and that the adversary could have earned a profit exceeding \ImprovementPumpAndArbitrage.

\begin{figure}[tb!]
    \centering
    \includegraphics[width=\textwidth]{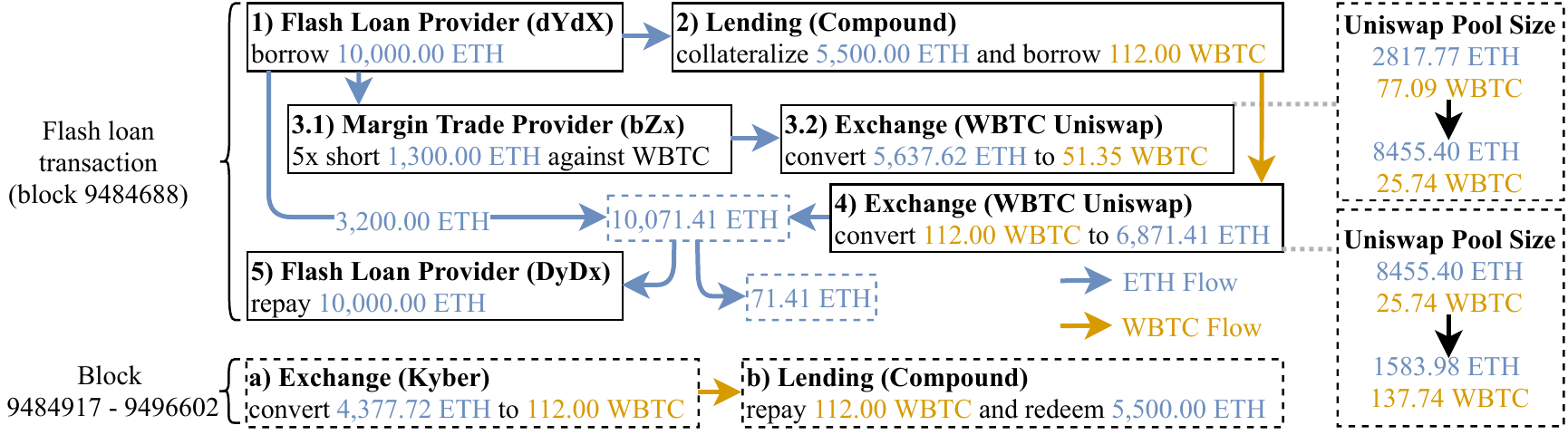}
    \caption{The pump attack and arbitrage. The attack consists of two parts, a flash loan and several loan redemption transactions.}
    \label{fig:attack_1}
\end{figure}

\point{Attack intuition}
The core of PA\&A is that the adversary pumps the price of ETH/WBTC on a constant product AMM DEX (Uniswap) with the leveraged funds of ETH in a margin trade. The adversary then purchases ETH at a ``cheaper'' price on the distorted DEX market (Uniswap) with the borrowed WBTC from a lending platform (Compound). As shown in Figure~\ref{fig:attack_1}, this attack mainly consists of two parts. For simplicity, we omit the conversion between ETH and WETH (the 1:1 convertible ERC20 version of ETH).

\point{Flash Loan (single transaction)}
The first part of the attack (cf.\ Figure~\ref{fig:attack_1}) consists of~$5$ steps within a single transaction. In step \step{1}, the adversary borrows a flash loan of~$10,000.00$~ETH from a flash loan provider~(dYdX). In step \step{2}, the adversarial trader collateralizes~$5,500.00$~ETH into a lending platform (Compound) to borrow~$112.00$~WBTC.
Note that the adversarial trader does not return the~$112.00$ WBTC within the flash loan. This means the adversarial trader takes the risk of a forced liquidation against the~$5,500.00$~ETH collateral if the price fluctuates. In steps~\step{3}, the trader provides~$1,300$~ETH to open a short position for ETH against WBTC (on bZx) with a~$5 \times$ leverage. Upon receiving this request, bZx transacts~$5,637.62$~ETH on an exchange (Uniswap) for only~$51.35$~WBTC (at~$109.79$ ETH/WBTC). Note that at the start of block~\block{9484688}, Uniswap has a total supply of~$2,817.77$ ETH and~$77.09$ WBTC (at~$36.55$ ETH/WBTC). The slippage of this transaction is significant with~$\frac{109.79 - 36.55}{36.55} = 200.38\%$. 
In step~\step{4}, the trader converts~$112.00$ WBTC borrowed from lending platform (Compound) to~$6,871.41$~ETH on the DEX (Uniswap) (at $61.35$ ETH/WBTC). 
We remark that the equity of the adversarial margin account is negative after the margin trading because of the significant price movement.
The pump attack could have been avoided if bZx checked the negative equity and reverted the transaction. At the time of the attack, this logic existed in the bZx contracts but was not invoked properly.
In step~\step{5}, the trader pays back the flash loan plus an interest of~$10^{-7}$~ETH. After the flash loan transaction (i.e., the first part of PA\&A), the trader gains~$71.41$~ETH, and has a debt of~$112$ WBTC over-collateralized by~$5,500$ ETH ($49.10$ ETH/WBTC). If the ETH/WBTC market price is below this loan exchange rate, the adversary can redeem the loan's collateral as follows.

\point{Loan redemption} 
The second part of the trade consists of two recurring steps, (step~\step{a}~-~\step{b}), between Ethereum block~\block{9484917} and \block{9496602}. Those transactions aim to redeem ETH by repaying the WBTC borrowed earlier (on Compound). To avoid slippage when purchasing WBTC, the trader executes the second part in small amounts over a period of two days on the DEX (Kyber, Uniswap). In total, the adversarial trader exchanged~$4,377.72$ ETH for~$112$ WBTC (at~$39.08$ ETH/WBTC) to redeem~$5,500.00$ ETH.

\point{Identifying the victim}
We investigate who of the participating entities is losing money. Note that in step \step{3} of Figure~\ref{fig:attack_1}, the short position (on bZx) borrows $5,637.62-1,300=4,337.62$ ETH from the lending provider (bZx), with~$1,300$ ETH collateral. Step \step{3} requires to purchase WBTC at a price of~$109.79$ ETH/WBTC, with both, the adversary's collateral and the pool funds of the liquidity provider. $109.79$ ETH/WBTC does not correspond to the market price of~$36.55$ ETH/WBTC prior to the attack, hence the liquidity provider overpays by nearly $3\times$ of the WBTC price.

\point{How much are the victims losing}
We now quantify the losses of the liquidity providers. The loan provider lose $4,337.62$ (ETH from loan providers) - $51.35$ (WBTC left in short position) $\times$ $39.08$ (market exchange rate ETH/WBTC) = $2,330.86$ ETH. The adversary gains~$5,500.00$ (ETH loan collateral in Compound) - $4,377.72$ (ETH spent to purchase WBTC) + $71.41$ (part 1) = $1,193.69$ ETH.

\point{More money is left on the table}
Due to the attack, Uniswap's price of ETH was reduced from~$36.55$ to~$11.50$ ETH/WBTC. This creates an arbitrage opportunity, where a trader can sell ETH against WBTC on Uniswap to synchronize the price. $1,233.79$ ETH would yield~$60.65$ WBTC, instead of~$33.76$ WBTC, realizing an arbitrage profit of $26.89$ WBTC ($286,035.04$ USD).

\subsection{Oracle Manipulation Attack}
We proceed to detail a second flash loan attack, which yields a profit of $2,381.41$ ETH (c.\ $634.9$k USD) within a single transaction (cf.\ \transaction{0x762881b07feb63c436dee38edd4ff1f7a74c33091e534af56c9f7d49b5ecac15}, on the~$18$th of February,~$2020$, at an ETH price of~$282.91$ USD/ETH) given a transaction fee of~$118.79$ USD. Before diving into the details, we cover additional background knowledge. We again show how the chosen attack parameters were sub-optimal and optimal parameters would yield a profit of \ImprovementOracleManipulation instead (cf.\ Section~\ref{sec:optimizing-existing-attacks2}).

\point{Price oracle}
One of the goals of the DeFi ecosystem is to not rely on trusted third parties. This premise holds both for asset custody as well as additional information, such as asset pricing. One common method to determine an asset price is hence to rely on the pricing information of an on-chain DEX (e.g., Uniswap). DEX prices, however, can be manipulated with flash loans.

\point{Attack intuition}
The core of this attack is an oracle manipulation using a flash loan, which lowers the price of sUSD/ETH. In a second step, the adversary benefits from this decreased sUSD/ETH price by borrowing ETH with sUSD as collateral.

\begin{figure}[tb!]
    \centering
    \includegraphics[width=\textwidth]{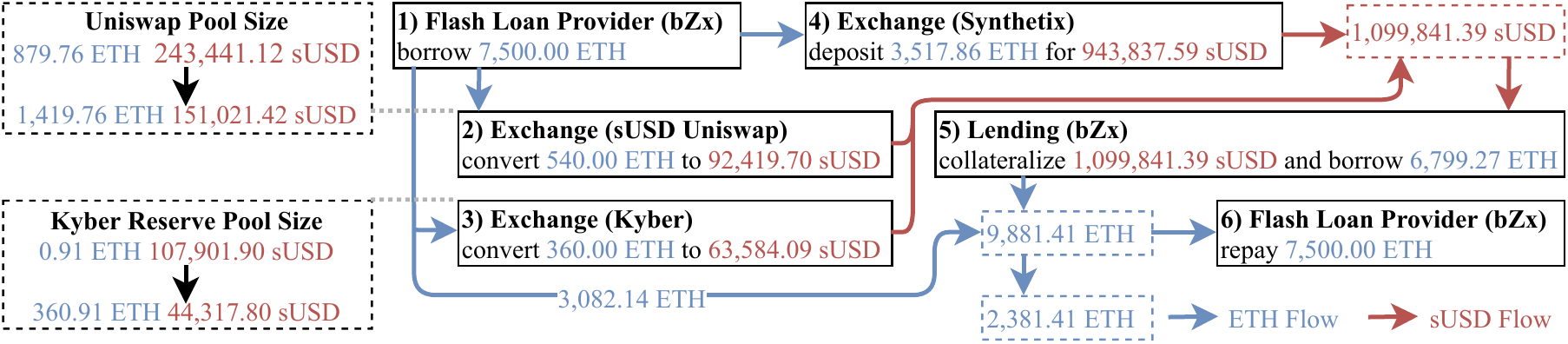}
    \caption{The oracle manipulation attack.}
    \label{fig:attack_2}
\end{figure}

\point{Adversarial oracle manipulation} 
We identify a total of~6 steps within this transaction (cf.\ Figure~\ref{fig:attack_2}). In step~\step{1}, the adversary borrows a flash loan of~$7,500.00$ ETH (on bZx). In the next three steps (\step{2},\step{3},\step{4}), the adversary converts a total of~$4,417.86$ ETH to~$1,099,841.39$ sUSD (at an average of~$248.95$ sUSD/ETH). 
The exchange rates in step \step{2} and \step{3} are ~$171.15$ and~$176.62$ sUSD/ETH respectively. These two steps decrease the sUSD/ETH price to $106.05$ sUSD/ETH on Uniswap and $108.44$ sUSD/ETH on Kyber Reserve, which are collectively used as a price oracle of the lending platform (bZx). Note that Uniswap is a constant product AMM, while Kyber Reserve is an AMM following a different formula (cf.\ Appendix~\ref{app:defimodels}). The trade on the third market (Synthetix) in step \step{4} is yet unaffected by the previous trades. The adversarial trader then collateralizes all the purchased sUSD ($1,099,841.39$) to borrow~$6,799.27$ ETH (at~$\frac{\text{exchange rate}}{\text{collateral factor}} = \text{max}(106.05, 108.44) \times 1.5 = 162.66$~sUSD/ETH on bZx). Now the adversary possesses~$6,799.27+3,082.14$ ETH and in the last step repays the flash loan amounting to~$7,500.00$ ETH. The adversary, therefore, generates a revenue of~$2,381.41$~ETH while only paying~$0.42$ ETH~($118.79$ USD) transaction fees.


\point{Identifying the victim}
The adversary distorted the price oracle (Uniswap and Kyber) from~$268.30$ sUSD/ETH to~$108.44$ sUSD/ETH, while other DeFi platforms remain unaffected at~$268.30$ sUSD/ETH. Similar to the pump attack and arbitrage, the lenders on bZx are the victims losing assets as a result of the distorted price oracle. The lender lost~$6,799.27$ ETH -~$1,099,841$ sUSD, which is estimated to be~$2,699.97$ ETH (at $268.30$ sUSD/ETH). The adversary gains $6,799.27$ (ETH from borrowing) - $3,517.86$ (ETH to purchase sUSD) - $360$ (ETH to purchase sUSD) - $540$ (ETH to purchase sUSD) = $2,381.41$ ETH.



\section{Optimizing DeFi Attacks}\label{sec:framework}
The atomicity of blockchain transactions guarantees the continuity of the action executions. When the initial state is deterministically known, this trait allows an adversary to predict the intermediate results precisely after each action execution and then to optimize the attacking outcome by adjusting action parameters.
In light of the complexity of optimizing DeFi attacks manually, we propose a \emph{constrained optimization framework} that is capable of optimizing the action parameters. We show, given a blockchain state and an attack vector composed of a series of DeFi actions, how an adversary can efficiently discover the optimal action parameters that maximize the resulting expected revenue.



\subsection{System and Threat Model}
The system considered is limited to one decentralized ledger which supports pseudo-Turing complete smart contracts (e.g., similar to the Ethereum Virtual Machine; state transitions can be reversed given certain conditions).

We assume the presence of one computationally bounded and economically rational adversary $\mathbb{A}$. $\mathbb{A}$  attempts to exploit the availability of flash loans for financial gain. While $\mathbb{A}$ is not required to provide its own collateral to perform the presented attacks, the adversary must be financially capable to pay transaction fees. The adversary may amass more capital which possibly could increase its impact and ROI.

\subsection{Parametrized Optimization Framework}\label{sec:parametrizedoptimization}

We start by modeling different components that may engage in a DeFi attack. To facilitate optimal parameter solving, we quantitatively formalize every endpoint provided by DeFi platforms as a state transition function $\mathsf{S}^\prime = \mathcal{T}(\mathsf{S}; p)$ with the constraints $\mathcal{C}(S; p)$, where $\mathsf{S}$ is the given state, $p$ are the parameters chosen by the adversary and $\mathsf{S}'$ is the output state. The state can represent, for example, the adversarial balance or any internal status of the DeFi platform, while the constraints are set by the execution requirements of the EVM (e.g., the Ether balance of an entity should never be a negative number) or the rules defined by the respective DeFi platform (e.g., a flash loan must be repaid before the transaction termination plus loan fees). When quantifying profits, we ignore the loan interest/fee payments and transaction fees, which are negligible in the present DeFi attacks. The constraints are enforced on the input parameters and output states to ensure that the optimizer yields valid parameters.

We define the balance state function $\mathcal{B}(\mathbb{E};\mathsf{X};\mathsf{S})$ to denote the balance of currency $\mathsf{X}$ held by entity $\mathbb{E}$ at a given state $\mathsf{S}$ and require Equation~\ref{eq:balance-constraints} to hold.

\begin{equation}\label{eq:balance-constraints}
 \forall(\mathbb{E},\mathsf{X},\mathsf{S}),\ \mathcal{B}(\mathbb{E};\mathsf{X};\mathsf{S}) \geq 0
\end{equation}
The mathematical DeFi models applied in this work are detailed in Appendix~\ref{app:defimodels}.

Our parametrized optimizer is designed to solve the optimal parameters that maximizes the revenue given an on-chain state, DeFi models and attack vector. An attack vector specifies the execution order of different endpoints across various DeFi platforms, depending on which we formalize a unidirectional chain of transition functions (cf.\ Equation~\ref{eq:transition_chain}).
\begin{equation}\label{eq:transition_chain}
    \mathsf{S}_{i}=\mathcal{T}_i(\mathsf{S}_{i-1}; p_{i})
\end{equation}

By nesting transition functions, we can obtain the cumulative state transition functions $\mathcal{ACC}_i(\mathsf{S}_0;p^{1:i})$ that satisfies Equation~\ref{eq:transition_acc}, where $p^{1:i}=(p_1, ..., p_i)$.
\begin{equation}\label{eq:transition_acc}
\begin{split}
    \mathsf{S}_i &= \mathcal{T}_i(\mathsf{S}_{i-1}; p_{i})=\mathcal{T}_i(\mathcal{T}_{i-1}(\mathsf{S}_{i-2};p_{i-1}); p_{i})\\
    &=\mathcal{T}_i(\mathcal{T}_{i-1}(... \mathcal{T}_{1}(\mathsf{S}_0,p_1)... ; p_{i-1});p_i)=\mathcal{ACC}_i(\mathsf{S}_0;p^{1:i})
\end{split}
\end{equation}
Therefore the constraints generated in each step can be expressed as Equation~\ref{eq:constraint_acc}.
\begin{equation}\label{eq:constraint_acc}
    \mathcal{C}_i(\mathsf{S}_i;p_i)\Longleftrightarrow\mathcal{C}_i(\mathcal{ACC}_i(\mathsf{S}_0;p^{1:i});p_i)
\end{equation}
We assume an attack vector composed of $N$ transition functions. The objective function can be calculated from the initial state $\mathsf{S}_0$ and the final state $\mathsf{S}_N$ (e.g., the increase of the adversarial balance).
\begin{equation}\label{eq:object_function}
    \mathcal{O}(\mathsf{S}_0;\mathsf{S}_N)\Longleftrightarrow\mathcal{O}(\mathsf{S}_0;\mathcal{ACC}(\mathsf{S}_0;p^{1:N}))
\end{equation}
Given the initial state $\mathsf{S}_0$, we formulate an attack vector into a constrained optimization problem with respect to all the parameters $p^{1:N}$ (cf.\ Equation~\ref{eq:constrained_optimization}).
\begin{equation}\label{eq:constrained_optimization}
    \begin{gathered}
    \text{maximize}\quad \mathcal{O}(\mathsf{S}_0;\mathcal{ACC}(\mathsf{S}_0;p^{1:N}))\\
    \text{s.t.}\quad\mathcal{C}_i(\mathcal{ACC}_i(\mathsf{S}_0;p^{1:i});p_i)\quad \forall i \in [1, N]
    \end{gathered}
\end{equation}

\section{Evaluation}\label{sec:evaluation}
In the following, we evaluate our parametrized optimization framework on the existing attacks described in Section~\ref{sec:postmortem}. 
We adopt the Sequential Least Squares Programming~(SLSQP) algorithm from SciPy\footnote{\url{https://www.scipy.org/}. We use the \texttt{minimize} function in the \texttt{optimize} package.} to solve the constructed optimization problems. Our framework is evaluated on a Ubuntu~18.04.2 machine with $16$ CPU cores and $32$ GB RAM.

\subsection{Optimizing the Pump Attack and Arbitrage}\label{sec:optimizing-existing-attacks1}




We first optimize the pump attack and arbitrage. Figure~\ref{fig:parametrization} summarizes the notations and the on-chain state when the attack was executed (i.e., $\mathsf{S}_0$). We use these blockchain records as the initial state in our evaluation. $\mathsf{X}$ and $\mathsf{Y}$ denote ETH and WBTC respectively. In the PA\&A attack vector, we intend to tune the following two parameters, \textit{(i)} $p_1$: the amount of $\mathsf{X}$ collateralized to borrow $\mathsf{Y}$ (cf.~step \circled{2} and \circled{3} in Figure~\ref{fig:attack_1}) and \textit{(ii)} $p_2$: the amount of $\mathsf{X}$ collateralized to short $\mathsf{Y}$ (cf.~step \circled{4} in Figure~\ref{fig:attack_1}). Following the methodology specified in Section~\ref{sec:parametrizedoptimization}, we derive the optimization problem and the corresponding constraints, which are presented in Figure~\ref{fig:contraints-arbitrage}. We detail the deriving procedure in Appendix~\ref{app:optimizingpumpandarbitrageattack}. We remark that there are five linear constraints and only one nonlinear constraint, which implies that the optimization can be solved efficiently. 

\begin{figure}[tb!]
\centering
\makebox[\textwidth]{%
    \begin{minipage}[b]{0.5\textwidth}
      \centering
      \scriptsize
      \renewcommand{\arraystretch}{1.3}
      \resizebox{1\textwidth}{!}{%
      \begin{tabular}{@{}|llr|@{}}
\hline
\bf Description & \bf Variable & \bf Value  \\
\hline\hline

Maximum Amount of ETH to flash loan      &     $v_{\mathsf{X}}$                & $10,000$\\\hline

Collateral Factor   & $\mathsf{cf}$                            & $0.75$                      \\
Collateralized Borrowing Exchange Rate       & $\mathsf{er}$                                & $36.48$                 \\
Maximum Amount of  WBTC to Borrow& $z_{\mathsf{Y}}$ & $155.70$       \\ \hline
Uniswap Reserved ETH        &           $u_\mathsf{X}(\mathsf{S}_0)$                    & $2,817.77$ \\
Uniswap Reserved WBTC       &           $u_\mathsf{Y}(\mathsf{S}_0)$        & $77.08$        \\
\hline
Over Collateral Ratio& $\mathsf{ocr}$                       & $1.153$                     \\
Leverage& $\ell$                                            & $5$                    \\
Maximum Amount of ETH to leverage &       $w_\mathsf{X}$                   &  $4,858.74$                    \\ \hline
Market Price of WBTC                      &         $\mathsf{p}_m$                 &              $39.08$        \\ 
\hline
\end{tabular}%
}
\captionsetup{width=0.95\textwidth}
\caption{Initial on-chain states of the PA\&A.}
\label{fig:parametrization}
\end{minipage}

\begin{minipage}[b]{0.5\textwidth}
\centering
\footnotesize
\renewcommand{\arraystretch}{1.3}
\renewcommand{\arraystretch}{1.2}
\resizebox{1\textwidth}{!}{%
\begin{tabular}{@{}|c|r|@{}}
\hline

\bf Objective   & \multirow{2}{*}{${u_\mathsf{X}(\mathsf{S}_0) + \frac{p_2 \times \ell}{\mathsf{ocr}}} - {u_\mathsf{X}(\mathsf{S}_4)} - p_2 - \frac{p_1\times \mathsf{cf} \times \mathsf{p}_m}{\mathsf{er}}$}                                            \\
\bf function    &                                                                  \\

\hline\hline
\multirow{6}{*}{\bf Constraints} 
                             & $p_1 \geq 0$, $p_2 \geq 0$\\
                             & $v_{\mathsf{X}} - p_0 - p_1 \geq 0$\\
                             & $z_{\mathsf{Y}}- \frac{p_1\times\mathsf{cf}}{\mathsf{er}}\geq 0$\\
                             & $w_{\mathsf{X}}+p_2-\frac{p_2\times \ell}{\mathsf{ocr}} \geq 0$\\
                             & $\mathsf{B}_0 + u_\mathsf{X}(\mathsf{S}_0) + \frac{p_2 \times \ell}{\mathsf{ocr}} - u_\mathsf{X}(\mathsf{S}_4) - p_1 - p_2 \geq 0$ \\
\hline
\end{tabular}%
}
\captionsetup{width=0.95\textwidth}
\caption{Generated PA\&A constraints. $u_\mathsf{X}(\mathsf{S}_4)$ is nonlinear with respect to~$p_1$ and~$p_2$.}
\label{fig:contraints-arbitrage}
    \end{minipage}
    }
\end{figure}

We repeated our experiment for $1,000$ times, the optimizer spent $6.1$ms on average converging to the optimum.
The optimizer provides a maximum revenue of~$2,778.94$ ETH when setting the parameters $(p_1; p_2)$ to $(2,470.08$; $1,456.23)$, while in the original attack the parameters $(5,500; 1,300)$ only yield $1,171.70$~ETH. Due to the ignorance of trading fees and precision differences, there is a minor discrepancy between the original attack revenue calculated with our model and the real revenue which is~$1,193.69$~ETH (cf.\ Section~\ref{sec:postmortem}).
This is a \ImprovementPumpAndArbitrage gain over the attack that took place, using the price of ETH at that time.
We experimentally validate the optimal PA\&A parameters by forking the Ethereum blockchain with Ganache~\cite{GanacheT10:online} at block~\block{9484687} (one block prior to the original attack transaction). We then implement the pump attack and arbitrage in solidity~v0.6.3. The revenue of the attack is divided into two parts: part one from the flash loan transaction, and part two which is a follow-up operation in later blocks (cf.\ Section~\ref{sec:postmortem}) to repay the loan. For simplicity, we chose to only validate the first part, abiding by the following methodology: \textit{(i)} We apply the parameter output of the parametrized optimizer, i.e., $(p_1; p_2) = (2,470.08;~1,456.23)$ to the adversarial validation smart contract. \textit{(ii)} Note that our model is an approximation of the real blockchain transition functions. Hence, due to the inaccuracy of our model, we cannot directly use the precise model output, but instead use the model output as a guide for a manual, trial, and error search. We find~$1,344$ is the maximum value of $p_2$ that allows the successful adversarial trade. \textit{(iii)} Given the new $p_2$ constraint, our optimizer outputs the new optimal parameters $(2,404;~1,344)$. \textit{(iv)} Our optimal adversarial trade yields a profit of~$1,958.01$ ETH on part one (as opposed to~$71.41$ ETH) and consumes a total of~$3.3$M gas.

\subsection{Optimizing the Oracle Manipulation Attack}\label{sec:optimizing-existing-attacks2}

In the oracle manipulation attack, we denote $\mathsf{X}$ as ETH and $\mathsf{Y}$ as sUSD, while the initial state variables are presented in Figure~\ref{fig:oraclemanipulationstate}. We assume that $\mathbb{A}$ owns zero balance of~$\mathsf{X}$ or~$\mathsf{Y}$. There are three parameters to optimize in this attack, \textit{(i)} $p_1$: the amount of $\mathsf{X}$ used to swap for $\mathsf{Y}$ in step 2); \textit{(ii)} $p_2$: the amount of $\mathsf{X}$ used to swap for $\mathsf{Y}$ in step 3); \textit{(iii)} $p_3$: the amount of $\mathsf{X}$ used to exchange for $\mathsf{Y}$ in step 4). We summarize the produced optimization problem and its constraints in Figure~\ref{fig:contraints-oracle}, of which five constraints are linear and the other two are nonlinear. We present the details in Appendix~\ref{app:optimizingtheoraclemanipulationattack}.


\begin{figure}[tb]
    \centering
    \makebox[\textwidth]{%
    \begin{minipage}[b]{0.5\textwidth}
      \centering
      \scriptsize
      \renewcommand{\arraystretch}{1.3}
      \resizebox{1\textwidth}{!}{%
      \begin{tabular}{@{}|llr|@{}}
\hline
\bf Description & \bf Variable & \bf Value  \\
\hline\hline

Maximum ETH to flash loan      &     $v_{\mathsf{X}}$                & $7,500$\\\hline

Uniswap Reserved ETH        &           $u_\mathsf{X}(\mathsf{S}_0)$                    & $879.757$ \\
Uniswap Reserved sUSD       &           $u_\mathsf{Y}(\mathsf{S}_0)$        & $243,441.12$        \\
\hline
Liquidity Rate& $\mathsf{lr}$                       &   $0.00252$               \\
Min.\ sUSD Price of Kyber Reserve & $\mathsf{minP}$                                            &    $0.0037$                \\
Max.\ sUSD Price of Kyber Reserve &       $\mathsf{maxP}$                   &         $0.0148$             \\
Inventory of ETH in Kyber Reserve & $k_{\mathsf{X}(\mathsf{S}_0)}$ & $0.90658$  \\\hline
Market Price of sUSD                   &         $\mathsf{p}_m$                 &              $0.00372719$        \\
Max.\ sUSD to Buy & $\mathsf{maxY}$  &   $943,837.59$ \\
\hline
Collateral Factor   & $\mathsf{cf}$                            & $0.667$                     \\
Max.\ ETH to Borrow& $z_{\mathsf{Y}}$ & $11,086.29$       \\ \hline
\end{tabular}%
}
\captionsetup{width=0.95\textwidth}
\caption{Initial on-chain states of the oracle manipulation attack.}
\label{fig:oraclemanipulationstate}
\end{minipage}

\begin{minipage}[b]{0.5\textwidth}
      \centering
      \footnotesize
      \renewcommand{\arraystretch}{1.3}
\renewcommand{\arraystretch}{1.2}
\resizebox{\textwidth}{!}{%
\begin{tabular}{@{}|c|r|@{}}
\hline
\bf Objective   & \multirow{2}{*}{$\mathcal{B}(\mathbb{A};\mathsf{Y};\mathsf{S}_4)\times \mathsf{cf} \times \mathsf{P}_\mathsf{Y}(\mathbb{M};\mathsf{S}_2) - p_1 -p_2 -p_3$}                                             \\
\bf function    &                                                                               \\
\hline\hline
\multirow{7}{*}{\bf Constraints} 
                             & $p_1 \geq 0$, $p_2 \geq 0$, $p_3 \geq 0$ \\
                             & $v_{\mathsf{X}} - p_1 - p_2 - p_3\geq 0$  \\
                        & $\mathsf{maxP} - \mathsf{minP} \times e^{\mathsf{lr}\times (k_\mathsf{X}(\mathsf{S}_0) + p_2)}\geq 0$\\
                        & $\mathsf{maxY} - \frac{p_3}{\mathsf{p}_m} \geq 0$\\
                        & $z_\mathsf{Y} - \mathcal{B}(\mathbb{A};\mathsf{Y};\mathsf{S}_4)\times \mathsf{cf} \times \mathsf{P}_\mathsf{Y}(\mathbb{M};\mathsf{S}_2) \geq 0$\\
\hline
\end{tabular}%
}
\captionsetup{width=0.95\textwidth}
\caption{Constraints generated for the oracle manipulation attack. $\mathcal{B}(\mathbb{A};\mathsf{Y};\mathsf{S}_4)$, $\mathsf{P}_\mathsf{Y}(\mathbb{M};\mathsf{S}_2)$ are nonlinear components with respect to $p_1$, $p_2$, $p_3$.}
\label{fig:contraints-oracle}
    \end{minipage}
    }
\end{figure}

We execute our optimizer~$1,000$ times, resulting in an average convergence time of~$12.9$ms. The optimizer discovers that setting $(p_1; p_2; p_3)$ to $(898.58$;$546.80$; $3,517.86)$ results in~$6,323.93$~ETH in profit for the adversary. This results in a gain of~\ImprovementOracleManipulation instead of~$634.9$k~USD. We fork the Ethereum blockchain with Ganache at block~\block{9504626} (one block prior to the original adversarial transaction) and again implement the attack in solidity~v0.6.3. We validate that executing the adversarial smart contract with parameters $(p_1; p_2; p_3) = (898.58$; $546.8$; $3,517.86)$ renders a profit of $6,262.28$ ETH, while the original attack parameters yield $2,381.41$ ETH. The attack consumes~$11.3$M gas (which fits within the current block gas limit of~$12.5$M gas, but wouldn't have fit in the block gas limit of February 2020). By analyzing the adversarial validation contract, we find that~$460$ is the maximum value of $p_2$ which reduces the gas consumption below $10$M gas. Similar to Section~\ref{sec:optimizing-existing-attacks1}, we add the new constraint to the optimizer, which then gives the optimal parameters~$(714.3$; $460$; $3,517.86)$. The augmented validation contract renders a profit of~$4,167.01$ ETH and consumes~$9.6$M gas.

\definecolor{reddot}{RGB}{216, 0, 115}
\definecolor{bluedot}{RGB}{27, 161, 226}

\begin{figure}[tb]
\makebox[\textwidth]{%
\begin{minipage}[b]{0.5\textwidth}
      \centering
      \includegraphics[width=1\textwidth]{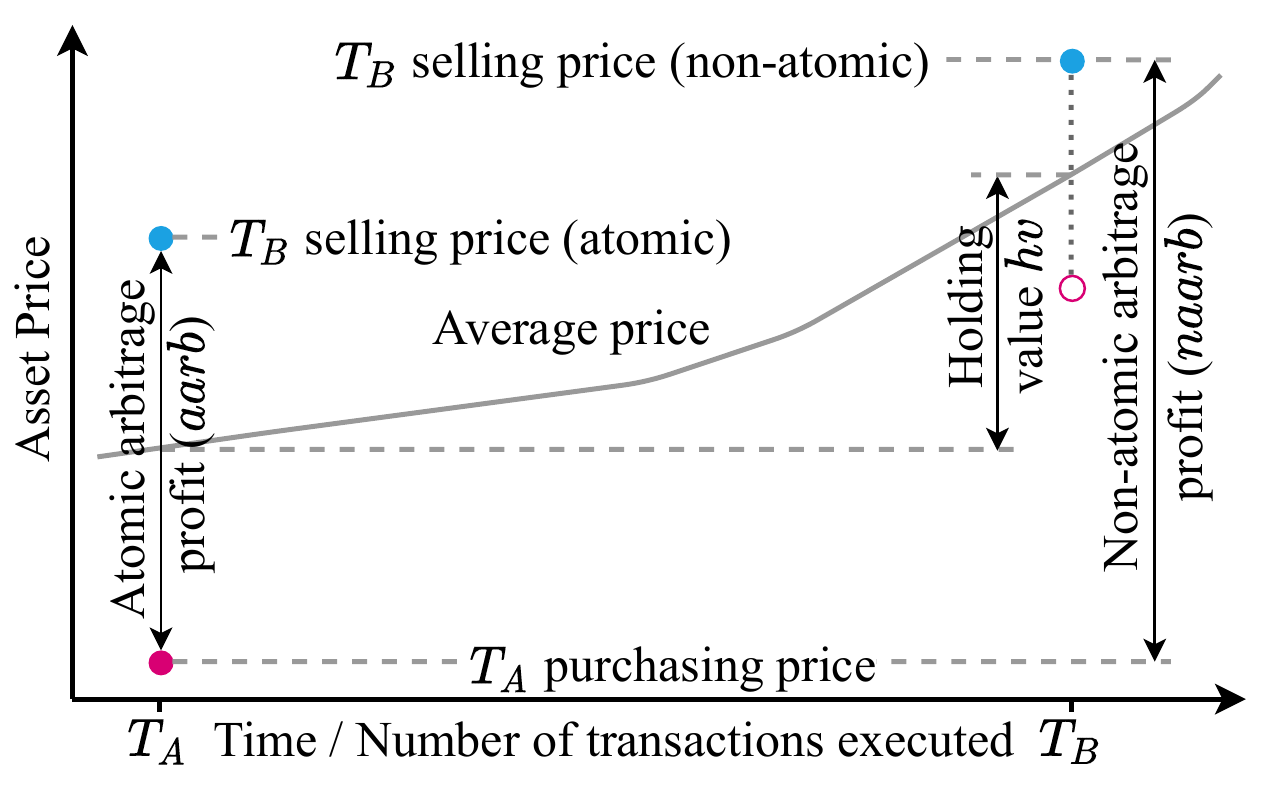}
      \captionsetup{width=0.95\textwidth}
      \caption{On the impact of transaction atomicity on arbitrage. The arbitrageur submits the first trade $T_A$, which aims to purchase an asset at a ``cheaper'' prices ({\color{reddot}$\bullet$}) and sell the asset on another exchange at a ``higher'' price ({\color{bluedot}$\bullet$}). In a non-atomic environment, $T_B$ is not immediately executed after $T_A$. The holding value is the in/decrease in price when holding the asset between $T_A$ and $T_B$.}
      \label{fig:aarb}
\end{minipage}
\begin{minipage}[b]{0.5\textwidth}
      \centering
      \includegraphics[width=0.95\textwidth]{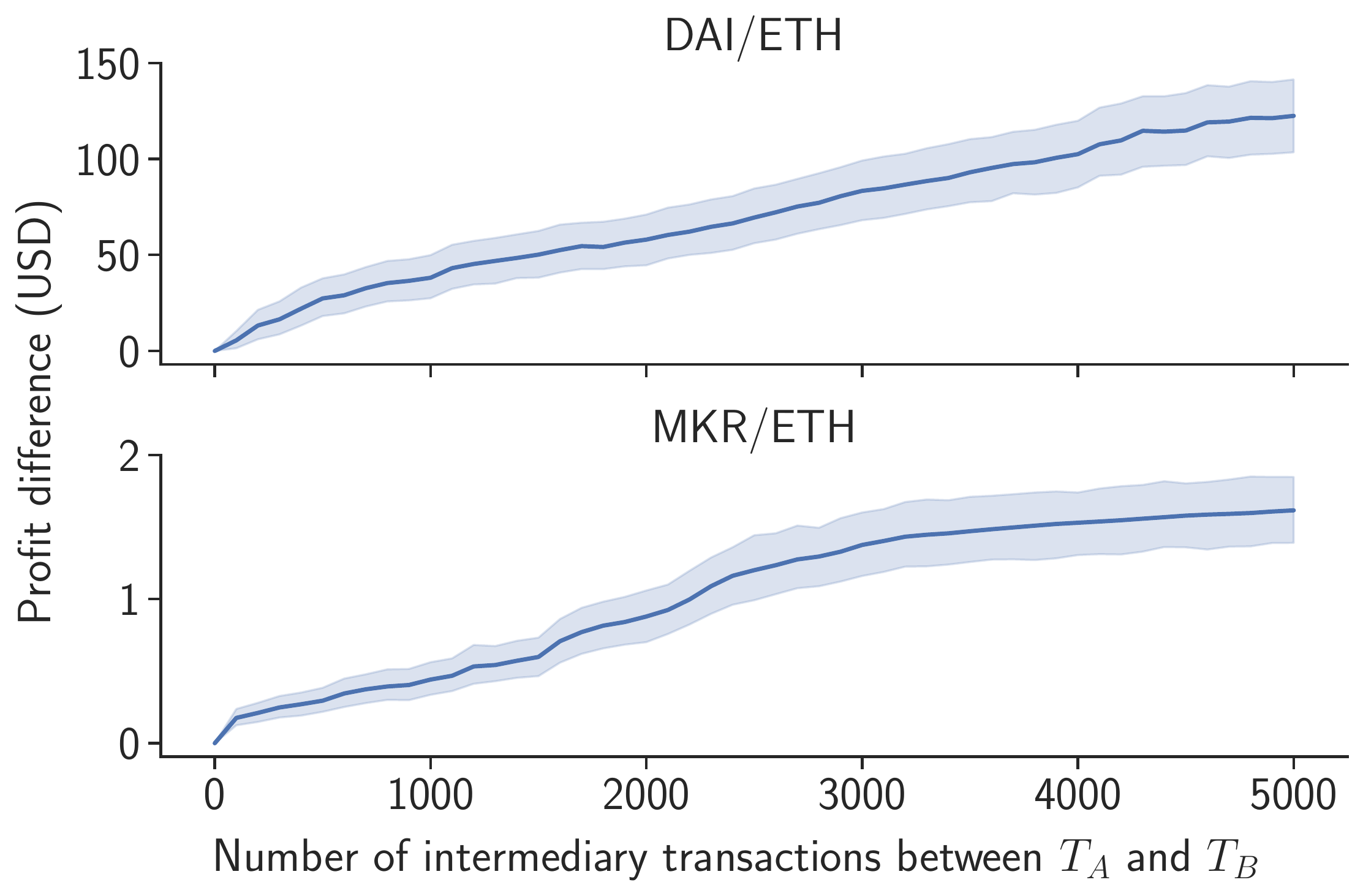}
      \captionsetup{width=0.95\textwidth}
      \caption{Simulated impact of intermediary transactions on arbitrage revenue. The average reward decreases by~$123.49 \pm 1375.32$~USD and~$1.77 \pm 10.59$~USD for the DAI/ETH and MKR/ETH markets respectively, at~$350$~USD/ETH, for~$5,000$ intermediary transactions. Note that we present the~$95\%$ bootstrap confidence interval of mean~\cite{diciccio1996bootstrap} for readability.}
      \label{fig:arbitrage_non_atomic}
\end{minipage}
}%
\end{figure}

\section{Implications of Transaction Atomicity}
\label{sec:atomicity}
In an atomic blockchain transaction, actions can be executed collectively in sequence, or fail collectively. Technically, operating DeFi actions in an atomic transaction is equivalent to acquiring a lock on all involved financial markets to ensure no other market agent can modify market states intermediately, and releasing the lock after executing all actions in their sequence.

To quantify objectively the impact of transaction atomicity (specifically, how the transaction atomicity impacts arbitrage profit), we proceed with the following methodology. We consider the arbitrages that involve two trades $T_A$ and $T_B$ to empirically compare the atomic and non-atomic arbitrages (cf. Figure~\ref{fig:aarb}). We define the atomic and non-atomic arbitrage profit as follows.
\point{Atomic arbitrage profit ($aarb$)} is defined as the gain of two atomically executed arbitrage trades $T_A$ and $T_B$ on exchange $A$ and $B$.
\point{Non-atomic arbitrage profit ($naarb$)} is defined as the arbitrage gain, if $T_A$ executes first, and $T_B$'s execution follows after $i$ intermediary transactions.

Conceptually, a non-atomic arbitrage requires the arbitrageur to lock assets for a short time (order of seconds/minutes). Those assets are exposed to price volatility. The arbitrageur can at times realize a gain, if the asset increases in value, but equally has the risk of losing value. A trader engaging in atomic arbitrage is not exposed to this volatility risk, which we denote as \emph{holding value}.
\point{Holding value ($hv$)} is defined as the change in the averaged price of the given asset pair on the two exchanges, which represents the asset value change during the non-atomic execution period.

We introduce holding value to neutralize the price volatility and can hence objectively quantify the financial advantage of atomic arbitrage. Given these variables, we define the \emph{profit difference} in Equation~\ref{eq:profit-difference}.
\begin{equation}\label{eq:profit-difference}
    \text{profit difference = $aarb$ - ($naarb$ - $hv$)}
\end{equation}

We simulate atomic and non-atomic based on~$6,398,992$ transactions we collect from the Ethereum mainnet (from block~\block{10276783} onwards). We insert~$0$ - $5,000$ blockchain transactions following the trade transaction $T_A$. Note that~$0$ intermediary transaction is equivalent to the atomic arbitrage. The insertion order follows the original execution order of these transactions, some of which may be irrelevant to the arbitrage.
We present the simulated profit difference in Figure~\ref{fig:arbitrage_non_atomic}. We observe that the average profit difference reaches~$123.49 \pm 1375.32$~USD and~$1.77 \pm 10.59$~USD for the DAI/ETH and MKR/ETH markets respectively when the number of intermediary transactions increases to~$5,000$.

\section{Discussion}\label{sec:discussion}
The current generation of DeFi had developed organically, without much scrutiny when it comes to financial security; it, therefore, presents an interesting security challenge to confront.  DeFi, on the one hand, welcomes innovation and the advent of new protocols, such as MakerDAO, Compound, and Uniswap. On the other hand, despite a great deal of effort spent on trying to secure smart contacts~\cite{luu2016making,jiang2018contractfuzzer,echidna2020,wustholz2019harvey,tsankov2018securify}, and to avoid various forms of market manipulation, etc.~\cite{mavroudis2019market,mavroudis2019libra,bentov2017tesseract}, there has been little-to-no effort to secure entire \emph{protocols}.

As such, DeFi protocols join the ecosystem, which leads to both exploits against protocols themselves as well as multi-step attacks that utilize several protocols such as the two attacks in Section~\ref{sec:postmortem}. In a certain poignant way, this highlights the fact the DeFi, lacking a central authority that would enforce a strong security posture, is ultimately vulnerable to a multitude of attacks by design. Flash loans are merely a mechanism that \emph{accelerates} these attacks. It does so by requiring no collateral (except for the minor gas costs), which is impossible in the traditional fiance due to regulations. In a certain way, flash loans democratize the attack, opening this strategy to the masses. 
As we anticipate in the earlier version of this paper, following the two analyzed attacks, economic attacks facilitated by flash loans become increasingly frequent, which have incurred a total loss of over $100$M~USD~\cite{HomePrev40:online}. 

\point{Determining what is malicious}
An interesting question remains whether we can qualify the use of flash loans
, as clearly malicious (or clearly benign). We believe this is a difficult question to answer and prefer to withhold the value judgment. The two attacks in Section~\ref{sec:postmortem} are clearly malicious: the PA\&A involves manipulating the WBTC/ETH price on Uniswap; the oracle manipulation attack involves price oracle by manipulatively lowering the price of ETH against sUSD on Kyber. However, the arbitrage mechanism, in general, is not malicious~---~it is merely a consequence of the decentralized nature of the DeFi ecosystem, where many exchanges and DEXs are allowed to exist without much coordination with each other. As such, arbitrage will continue to exist as a phenomenon, with good and bad consequences. 
Despite the lack of absolute distinction between flash loan attacks and legitimate applications of flash loans, we attempt to summarize two characteristics that appear to apply to malicious flash loan attacks: \emph{(i)} the attacker benefits from a distorted state created artificially in the flash loan transaction (e.g., the pumped market in the PA\&A and the manipulated oracle price); \emph{(ii)} the attacker's profit causes the loss of other market participants (e.g., the liquidity providers in the two analyzed attacks in Section~\ref{sec:postmortem}).

We extend our discussion in Appendix~\ref{app:extended-discussion}.

\vspace{-1pt}
\section{Conclusion}\label{sec:conclusion}
This paper presents an exploration of the impact of transaction atomicity and the flash loan mechanism on the Ethereum network. While proposed as a clever mechanism within DeFi, flash loans are starting to be used as financial attack vectors to effectively pull money in the form of cryptocurrency out of DeFi.
In this paper, we analyze existing flash loan-based attacks in detail and then proceed to propose optimizations that significantly improve the ROI of these attacks.
Specifically, we are able to show how two previously executed attacks can be ``boosted'' to result in a revenue of~\ImprovementPumpAndArbitrage and~\ImprovementOracleManipulation, respectively, which is a boost of~\RatioPumpAndArbitrage and~\RatioOracleManipulation, respectively.

\section*{Acknowledgments}
We thank the anonymous reviewers and Johannes Krupp for providing valuable comments and helpful feedback that significantly strengthened the paper. We are moreover grateful to the Lucerne University of Applied Sciences and Arts for generously supporting Kaihua Qin's Ph.D.

\bibliographystyle{splncs04}
\bibliography{references.bib}

\appendix

\section{Classifying Flash Loan Use Cases}\label{app:classifyingflashloanusecases}
In Figure~\ref{fig:usage}, we present the DeFi platforms that use a total of $5,615$ Aave flash loan transactions\footnote{We collect in total $5,616$ flash loans with one transaction performing two flash loans.} between the~8th of January,~2020 and the~20th of September,~2020. We find that more than $30$\% of the flash loans are interacting with Kyber, MakerDAO, and Uniswap. Compound and MakerDAO accumulate~$433.81$M~USD flash loans which occupy $90$\% of the total flash loan amount. On average, a flash transaction uses $1.43$M gas, while the most complex one consumes $6.3$M gas.
\begin{figure}[htb!]
\centering
\scriptsize
\renewcommand{\arraystretch}{1.4}
\setlength{\tabcolsep}{2pt}
\resizebox{\textwidth}{!}{%
\begin{tabular}{|llll|}
\hline
\bf DeFi Platforms &  \bf Transactions &  \bf Amount (USD) &  \bf Mean gas \\
\hline\hline
Kyber, MakerDAO, Uniswap & $1826$ & $6.91$M & $1.64$M$\pm$$465.69$k \\
Kyber, MakerDAO, OasisDEX, Uniswap & $817$ & $6.75$M & $1.38$M$\pm$$324.09$k \\
Compound, MakerDAO & $320$ & $433.81$M & $1.49$M$\pm$$333.16$k \\
0x, Kyber, MakerDAO, Uniswap & $231$ & $888.17$k & $1.76$M$\pm$$595.93$k \\
Compound & $228$ & $5.98$M & $1.22$M$\pm$$501.97$k \\
0x, Compound, Curve, MakerDAO & $168$ & $115.82$k & $1.31$M$\pm$$603.77$k \\
0x, Kyber, MakerDAO, OasisDEX, Uniswap & $153$ & $2.12$M & $1.80$M$\pm$$432.11$k \\
Compound, Curve & $143$ & $1.75$M & $2.06$M$\pm$$281.84$k \\
MakerDAO & $122$ & $8.86$M & $934.39$k$\pm$$230.73$k \\
0x, Compound, Curve & $103$ & $103.00$k & $1.27$M$\pm$$249.15$k \\
Compound, MakerDAO, Uniswap & $93$ & $120.18$k & $1.31$M$\pm$$314.83$k \\
Kyber, Uniswap & $92$ & $80.54$k & $985.68$k$\pm$$711.43$k \\
0x, MakerDAO & $87$ & $1.70$M & $1.18$M$\pm$$120.70$k \\
Bancor, Compound, Kyber, MakerDAO, Uniswap & $77$ & $8.45$k & $2.14$M$\pm$$705.27$k \\
0x, Uniswap & $68$ & $32.97$k & $694.76$k$\pm$$129.58$k \\
MakerDAO, Uniswap & $68$ & $40.83$k & $1.01$M$\pm$$254.51$k \\
0x, OasisDEX & $57$ & $23.79$k & $716.40$k$\pm$$132.51$k \\
Kyber, MakerDAO & $53$ & $437.65$k & $2.06$M$\pm$$641.44$k \\
0x, Kyber, MakerDAO & $42$ & $639.36$k & $1.78$M$\pm$$352.44$k \\
Compound, Kyber, MakerDAO, Uniswap & $37$ & $185.30$k & $2.72$M$\pm$$740.48$k \\
0x, Kyber, Uniswap & $30$ & $23.81$k & $1.30$M$\pm$$285.27$k \\
Bancor, Compound, Kyber, MakerDAO, OasisDEX, Uniswap & $30$ & $13.46$k & $2.05$M$\pm$$666.87$k \\
Compound, Uniswap & $29$ & $45.58$k & $1.14$M$\pm$$293.59$k \\
MakerDAO, OasisDEX & $27$ & $114.31$k & $823.62$k$\pm$$139.90$k \\
Uniswap & $25$ & $56.34$k & $672.12$k$\pm$$404.84$k \\
0x, Compound, MakerDAO & $22$ & $88.57$k & $1.81$M$\pm$$274.23$k \\
Kyber & $21$ & $41.73$k & $803.54$k$\pm$$207.92$k \\
Compound, Curve, MakerDAO & $20$ & $3.10$M & $1.93$M$\pm$$665.87$k \\
Compound, Kyber, Uniswap & $13$ & $18.04$k & $1.82$M$\pm$$430.46$k \\
0x, Kyber, OasisDEX, Uniswap & $13$ & $11.99$k & $1.42$M$\pm$$291.46$k \\
0x, OasisDEX, Uniswap & $12$ & $15.68$k & $789.94$k$\pm$$193.06$k \\
Compound, Kyber, MakerDAO, OasisDEX, Uniswap & $11$ & $63.12$k & $3.20$M$\pm$$893.03$k \\
0x & $9$ & $8.48$k & $590.03$k$\pm$$111.78$k \\
Kyber, OasisDEX, Uniswap & $8$ & $42.55$k & $858.12$k$\pm$$255.44$k \\
0x, Compound, Curve, Kyber, MakerDAO, Uniswap & $7$ & $6.98$k & $1.87$M$\pm$$301.64$k \\
Kyber, MakerDAO, OasisDEX & $6$ & $130.31$k & $1.84$M$\pm$$512.57$k \\
0x, Compound, MakerDAO, Uniswap & $5$ & $2.64$k & $2.02$M$\pm$$149.59$k \\
Bancor, Compound, Kyber, Uniswap & $5$ & $564.52$ & $3.83$M$\pm$$1.50$M \\
Others & $537$ & $6.87$M & $670.22$k$\pm$ $568.05$k \\
\hline
Total & $5,615$ & $481.20$M & $1.43$M$\pm$ $605.97$k \\
\hline
\end{tabular}%
}
\caption{Classifying the usage of flash loans in the wild, based on an analysis of transactions between the~8th of January,~2020 and the~20th of September,~2020 on Aave~\cite{aave}. \emph{Others} include the platform combinations that appear less than five times and the ones of which the owner platforms are unknown to us. The total amount is calculated at the price~-- DAI (\$$1$); ETH (\$$350$); USDC (\$$1$); BAT (\$$0.2$); WBTC (\$$10,000$); ZRX (\$$0.3$); MKR (\$$500$); LINK (\$$10$); USDT (\$$1$); REP (\$$15$), KNC (\$$1.5$), LEND  (\$$0.5$), sUSD (\$$1$).}
\label{fig:usage}
\end{figure}

\section{Flash Loan Use Cases}\label{sec:flash-use-cases}


    


\subsection{Wash Trading}
The trading volume of an asset is a metric indicating its popularity. The most popular assets therefore are supposed to be traded the most --- e.g., Bitcoin to date enjoys the highest trading volume (reported up to $50$T USD per day) of all cryptocurrencies.

\begin{figure}[htb!]
  \begin{minipage}[c]{0.65\columnwidth}
    \includegraphics[width=\textwidth]{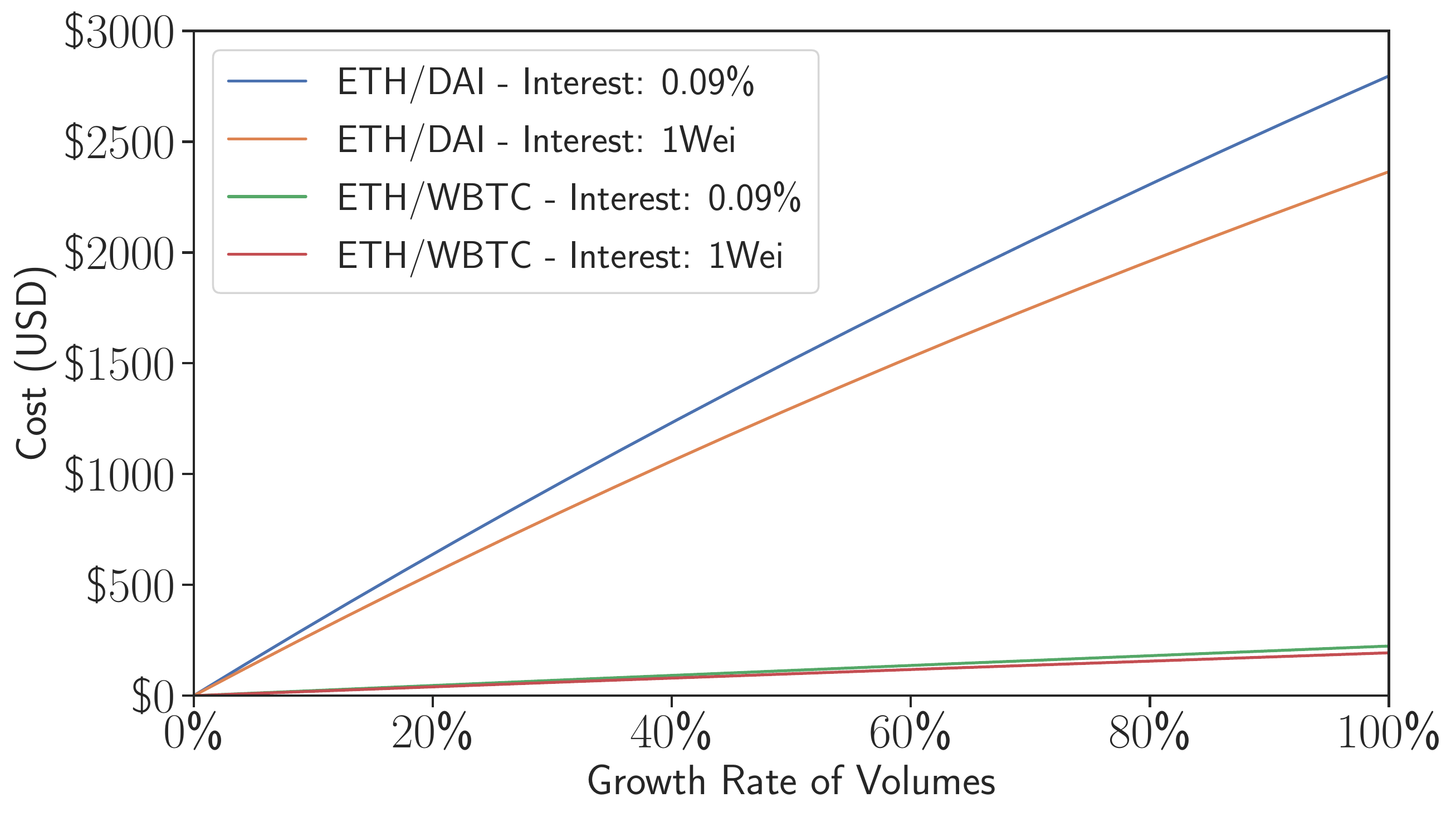}
  \end{minipage}\hfill
  \begin{minipage}[c]{0.35\columnwidth}
    \caption{Wash trading cost on two Uniswap markets with flash loans costing~$0.09\%$ (Aave) and a constant of 1 Wei (dYdX) respectively. The 24-hour volumes of ETH/DAI and ETH/WBTC market were~$963,786$ USD and~$67,690$ USD respectively (1st of March, 2020).
    } \label{fig:wash_trading_cost}
  \end{minipage}
\end{figure}

Malicious exchanges or traders can mislead other traders by artificially inflating the trading volume of an asset. In September~2019,~$73$ out of the top~$100$ exchanges on Coinmarketcap~\cite{coinmarketcap} were wash trading over~$90$\% of their volumes~\cite{BTIMarke7:online}. In centralized exchanges, operators can easily and freely create fake trades in the backend, while decentralized exchanges settle trades on-chain. Wash trading on DEX thus requires wash traders to hold and use real assets. Flash loans can remove this ``obstacle'' and wash trading costs are then reduced to the flash loan interest, trading fees, and (blockchain) transaction fees, e.g., gas. A wash trading endeavor to increase the~24-hour volume by~50\% on the ETH/DAI market of Uniswap would for instance cost about~$1,298$ USD (cf.\ Figure~\ref{fig:wash_trading_cost}). We visualize in Figure~\ref{fig:wash_trading_cost} the required cost to create fake volumes in two Uniswap markets. At the time of writing, the transaction fee amounts to~$0.01$ USD, the flash loan interests range from a constant~$1$ Wei (on dYdX) to~$0.09$\% (on Aave), and exchange fees are about~$0.3$\% (on Uniswap).


\point{Wash trading example} 
On March~2nd,~2020, a flash loan of~$0.01$ ETH borrowed from dYdX performed two back-and-forth trades (first converted~$0.01$ ETH to~$122.1898$ LOOM and then converted~$122.1898$ LOOM back to~$0.0099$ ETH) on Uniswap ETH/LOOM market (cf.\ \transaction{0xf65b384ebe2b7bf1e7bd06adf0daac0413defeed42fd2cc72a75385a200e1544}). The 24-hour trading volume of the ETH/LOOM market increased by~$25.8$\% (from~$17.71$ USD to~$22.28$ USD) as a result of the two trades.

\subsection{Collateral Swapping}
We classify DeFi platforms that rely on users providing cryptocurrencies~\cite{dydx,aave,makerdao} as follows: (i) a DeFi system where a new asset is minted and backed-up with user-provided collateral (e.g., MakerDAO's DAI or SAI~\cite{makerdao}) and (ii) a DeFi system where long-term loans are offered and assets are aggregated within liquidity pools (e.g., margin trading~\cite{bZxAProt64:online} or long term loans~\cite{aave}). Once a collateral position is opened, DeFi platforms store the collateral assets in a vault until the new/borrowed asset are destroyed/returned. Because cryptocurrency prices fluctuate, this asset lock-in bears a currency risk. With flash loans, it is possible to replace the collateral asset with another asset, even if a user does not possess sufficient funds to destroy/return the new/borrowed asset. A user can close an existing collateral position with borrowed funds, and then immediately open a new collateral position using a different asset.

\point{Collateral swapping example}
On February~$20th$,~$2020$, a flash loan borrowed~$20.00$ DAI (from Aave) to perform a collateral swap (on MakerDAO), cf.\ \transaction{0x5d5bbfe0b666631916adb8a56821b204d97e75e2a852945ac7396a82e207e0ca}. Before this transaction, the transaction sender used~$0.18$ WETH as collateral for instantiating~$20.00$ DAI (on MakerDAO). The transaction sender first withdraws all WETH using the~$20.00$ DAI flash loan, then converts~$0.18$ WETH for~$178.08$ BAT (using Uniswap). Finally the user creates~$20.03$ DAI using BAT as collateral, and pays back~$20.02$ DAI (with a fee to Aave). This transaction converts the collateral from WETH to BAT and the user gained~$0.01$ DAI, with an estimated gas fee of~$0.86$ USD.

\subsection{Flash Minting}
Cryptocurrency assets are commonly known as either inflationary (further units of an asset can be mined) or deflationary (the total number of units of an asset are finite). Flash minting is an idea to allow an instantaneous minting of an arbitrary amount of an asset~---~the newly-mined units exist only during one transaction. It is yet unclear where this idea might be applicable to, the minted assets could momentarily increase liquidity.

\point{Flash minting example}
A flash mint function (cf.\ Figure~\ref{lst:flashmint}) can be integrated into an ERC20 token, to mint an arbitrary number of coins within a transaction only. Before the transaction terminates, the minted coins will be burned. If the available amount of coins to be burned by the end of the transaction is less than those that were minted, the transaction is reverted (i.e., not executed). An example ERC20 flash minting code could take the following form (cf.\ \address{0x09b4c8200f0cb51e6d44a1974a1bc07336b9f47f}):

\begin{figure}[bt]
    \centering
\begin{lstlisting}[language=JavaScript,frame=single, basicstyle=\scriptsize]
contract FlashMintableCoin is ERC20 { [...]
    function flashMint(uint256 amount) {
        // mint coins and transfer them 
        mint(msg.sender, amount);
        // borrower uses the loan
        Borrower(msg.sender).execute(amount);
        // reverts if not have enough to burn
        burn(msg.sender, amount);
}}
\end{lstlisting}
    \caption{Flash mint example.}
    \label{lst:flashmint}
\end{figure}

\section{DeFi Models}\label{app:defimodels}
In the following, we detail the quantitative DeFi models applied in this work. Note that we do not include all the states involved in the DeFi attacks but only those relevant to the constrained optimization.

\point{Flash loan}
We assume a flash loan platform $\mathbb{F}$ with $z_\mathsf{X}$ amount of asset $\mathsf{X}$, which the adversary $\mathbb{A}$ can borrow. The required interest to borrow $b$ of $\mathsf{X}$ is represented by $\operatorname{interest}(b)$.

\states In a flash loan, the state is represented by the balance of $\mathbb{A}$, i.e., $\mathcal{B}(\mathbb{A};\mathsf{X};\mathsf{S})$.
\transitions We define the transition functions of \Loan\ in Equation~\ref{eq:loan_transition_functions_constraints} and \Repay\ in Equation~\ref{eq:repay_transition_functions_constraints}, where the parameter $b_{\mathsf{X}}$ denotes the loaned amount.
\begin{equation}\label{eq:loan_transition_functions_constraints}
\begin{gathered}
    \mathcal{B}(\mathbb{A};\mathsf{X};\mathsf{S}') = \mathcal{B}(\mathbb{A};\mathsf{X};\mathsf{S}) + b_{\mathsf{X}}\\
    \text{s.t.} \quad z_{\mathsf{X}} - b_{\mathsf{X}} \geq 0
\end{gathered}
\end{equation}

\begin{equation}\label{eq:repay_transition_functions_constraints}
\begin{gathered}
    \mathcal{B}(\mathbb{A};\mathsf{X};\mathsf{S}') = \mathcal{B}(\mathbb{A};\mathsf{X};\mathsf{S}) - b_{\mathsf{X}} - \operatorname{interest}(b_{\mathsf{X}})\\
    \text{s.t.} \quad \mathcal{B}(\mathbb{A};\mathsf{X};\mathsf{S}) - b_{\mathsf{X}} - \operatorname{interest}(b_{\mathsf{X}}) \geq 0
\end{gathered}
\end{equation}

\point{Fixed price trading} We define the endpoint \SellXY\ that allows the adversary $\mathbb{A}$ to trade $q_\mathsf{X}$ amount of $\mathsf{X}$
for $\mathsf{Y}$ at a fixed price $\mathsf{p}_m$. $\mathsf{maxY}$ is the maximum amount of $\mathsf{Y}$ available for trading.

\states We consider the following state variables:
\begin{itemize}
    \item Balance of asset $\mathsf{X}$ held by $\mathbb{A}$: $\mathcal{B}(\mathbb{A};\mathsf{X};\mathsf{S})$
    \item Balance of asset $\mathsf{Y}$ held by $\mathbb{A}$: $\mathcal{B}(\mathbb{A};\mathsf{Y};\mathsf{S})$
\end{itemize}

\transitions Transition functions of \SellXY\ are defined in Equation~\ref{eq:fixedpricetrading}.
\begin{equation}\label{eq:fixedpricetrading}
\begin{gathered}
\mathcal{B}(\mathbb{A};\mathsf{X};\mathsf{S}') = \mathcal{B}(\mathbb{A};\mathsf{X};\mathsf{S}) - q_\mathsf{X}\\
\mathcal{B}(\mathbb{A};\mathsf{Y};\mathsf{S}') = \mathcal{B}(\mathbb{A};\mathsf{Y};\mathsf{S}) + \frac{q_\mathsf{X}}{\mathsf{p}_m}\\
\text{s.t.}\quad \mathcal{B}(\mathbb{A};\mathsf{X};\mathsf{S}) - q_\mathsf{X} \geq 0\\
\mathsf{maxY} - \frac{q_\mathsf{X}}{\mathsf{p}_m} \geq 0
\end{gathered}
\end{equation}

\point{Constant product automated market maker}
The constant product AMM is with a market share of 77\% among the AMM DEX, the most common AMM model in the current DeFi ecosystem~\cite{UniswapH70:online}. We denote by $\mathbb{M}$ an AMM instance with trading pair $\mathsf{X}/\mathsf{Y}$ and exchange fee rate $\mathsf{f}$.\\
\states We consider the following states variables that can be modified in an AMM state transition.
\begin{itemize}
    \item Amount of $\mathsf{X}$ in AMM liquidity pool: $u_\mathsf{X}(\mathsf{S})$, which equals to $\mathcal{B}(\mathbb{M};\mathsf{X};\mathsf{S})$
    \item Amount of $\mathsf{Y}$ in AMM liquidity pool: $u_\mathsf{Y}(\mathsf{S})$, which equals to $\mathcal{B}(\mathbb{M};\mathsf{Y};\mathsf{S})$
    \item Balance of $\mathsf{X}$ held by $\mathbb{A}$: $\mathcal{B}(\mathbb{A};\mathsf{X};\mathsf{S})$
    \item Balance of $\mathsf{Y}$ held by $\mathbb{A}$: $\mathcal{B}(\mathbb{A};\mathsf{Y};\mathsf{S})$
\end{itemize}
\transitions Among the endpoints of $\mathbb{M}$, we focus on \SwapXY\ and \SwapYX, which are the relevant endpoints for the DeFi attacks discussed within this work. $p_{\mathsf{X}}$\ is a parameter that represents the amount of $\mathsf{X}$ the adversary intends to trade. $\mathbb{A}$ inputs $p_\mathsf{X}$ amount of $\mathsf{X}$ in AMM liquidity pool and receives $o_\mathsf{Y}$ amount of $\mathsf{Y}$ as output.
The constant product rule~\cite{UniswapH70:online} requires that Equation~\ref{eq:constant-amm} holds.
\begin{equation}\label{eq:constant-amm}
    u_\mathsf{X}(\mathsf{S})\times u_\mathsf{Y}(\mathsf{S}) = \left(u_\mathsf{X}(\mathsf{S}) + (1-\mathsf{f}) p_\mathsf{X}\right)\times \left(u_\mathsf{Y}(\mathsf{S})-o_\mathsf{Y}\right)
\end{equation}
We define the transition functions and constraints of \SwapXY\ in Equation~\ref{eq:amm_transition_functions_constraints} (analogously for \SwapYX\ ).
\begin{equation}
\label{eq:amm_transition_functions_constraints}
\begin{gathered}
    \mathcal{B}(\mathbb{A};\mathsf{X};\mathsf{S'}) = \mathcal{B}(\mathbb{A};\mathsf{X};\mathsf{S}) - p_{\mathsf{X}}\\
    \mathcal{B}(\mathbb{A};\mathsf{Y};\mathsf{S'}) = \mathcal{B}(\mathbb{A};\mathsf{Y};\mathsf{S}) + o_{\mathsf{Y}}\\
    u_\mathsf{X}(\mathsf{S}') = u_\mathsf{X}(\mathsf{S}) + p_{\mathsf{X}}\\
    u_\mathsf{Y}(\mathsf{S}') = u_\mathsf{Y}(\mathsf{S}) - o_{\mathsf{Y}}\\
    \text{where} \quad o_{\mathsf{Y}} = \frac{p_{\mathsf{X}}\times(1-\mathsf{f})\times u_\mathsf{Y}(\mathsf{S})}{u_\mathsf{X}(\mathsf{S}) + p_{\mathsf{X}}\times (1-\mathsf{f})}\\
    \text{s.t.} \quad \mathcal{B}(\mathbb{M};\mathsf{X};\mathsf{S}) - p_{\mathsf{X}} \geq 0
\end{gathered}
\end{equation}

Because an AMM DEX $\mathbb{M}$ transparently exposes all price transitions on-chain, it can be used as a price oracle by the other DeFi platforms. The price of $\mathsf{Y}$ with respect to $\mathsf{X}$ given by $\mathbb{M}$ at state $\mathsf{S}$ is defined in Equation~\ref{eq:ammprice}.
\begin{equation}\label{eq:ammprice}
    \mathsf{p}_\mathsf{Y}(\mathbb{M};\mathsf{S}) = \frac{u_\mathsf{X}(\mathsf{S})}{u_\mathsf{Y}(\mathsf{S})}
\end{equation}

\point{Automated price reserve}
The automated price reserve is another type of AMM that automatically calculates the exchange price depending on the assets held in inventory. We denote a reserve holding the asset pair $\mathsf{X}/\mathsf{Y}$ with $\mathbb{R}$. A minimum price $\mathsf{minP}$ and a maximum price $\mathsf{maxP}$ is set when initiating $\mathbb{R}$. $\mathbb{R}$ relies on a liquidity ratio parameter $\mathsf{lr}$ to calculate the asset price. We assume that $\mathbb{R}$ holds $k_\mathsf{X}(\mathsf{S})$ amount of $\mathsf{X}$ at state $\mathsf{S}$. We define the price of $\mathsf{Y}$ in Equation~\ref{eq:apsprice}.
\begin{equation}\label{eq:apsprice}
    \mathsf{P}_\mathsf{Y}(\mathbb{R};\mathsf{S})=\mathsf{minP}\times e^{\mathsf{lr}\times k_\mathsf{X}(\mathsf{S})}
\end{equation}
The endpoint \ConvertXY\ provided by $\mathbb{R}$ allows the adversary $\mathbb{A}$ to exchange $\mathsf{X}$ for $\mathsf{Y}$.

\states We consider the following state variables:
\begin{itemize}
     \item The inventory of $\mathsf{X}$ in the reserve: $k_\mathsf{X}(\mathsf{S})$, which equals to $\mathcal{B}(\mathbb{R};\mathsf{X};\mathsf{S})$
    \item Balance of $\mathsf{X}$ held by $\mathbb{A}$: $\mathcal{B}(\mathbb{A};\mathsf{X};\mathsf{S})$
    \item Balance of $\mathsf{Y}$ held by $\mathbb{A}$: $\mathcal{B}(\mathbb{A};\mathsf{Y};\mathsf{S})$
\end{itemize}

\transitions We denote as $h_\mathsf{X}$ the amount of $\mathsf{X}$ that $\mathbb{A}$ inputs in the exchange to trade against $\mathsf{Y}$. The exchange output amount of $\mathsf{Y}$ is calculated by the following formulation.
\begin{equation*}
j_\mathsf{Y} = \frac{e^{-\mathsf{lr}\times h_\mathsf{X}} - 1}{\mathsf{lr}\times \mathsf{P}_\mathsf{Y}(\mathbb{R};\mathsf{S})}
\end{equation*}

We define the transition functions within Equation~\ref{eq:apr_transitions}.
\begin{equation}\label{eq:apr_transitions}
\begin{gathered}
k_\mathsf{X}(\mathsf{S}') = k_\mathsf{X}(\mathsf{S}) + h_\mathsf{X}\\
\mathcal{B}(\mathbb{A};\mathsf{X};\mathsf{S}') = \mathcal{B}(\mathbb{A};\mathsf{X};\mathsf{S}) - h_\mathsf{X}\\
\mathcal{B}(\mathbb{A};\mathsf{Y};\mathsf{S}') =
\mathcal{B}(\mathbb{A};\mathsf{Y};\mathsf{S}) + j_\mathsf{Y}\\
\text{where} \quad j_\mathsf{Y} = \frac{e^{-\mathsf{lr}\times h_\mathsf{X}} - 1}{\mathsf{lr}\times \mathsf{P}_\mathsf{Y}(\mathbb{R};\mathsf{S})}\\
\text{s.t.} \quad \mathcal{B}(\mathbb{A};\mathsf{X};\mathsf{S}) - h_\mathsf{X} \geq 0\\
\mathsf{P}_\mathsf{Y}(\mathbb{R};\mathsf{S}') - \mathsf{minP}\geq 0\\
\mathsf{maxP}-\mathsf{P}_\mathsf{Y}(\mathbb{R};\mathsf{S}')\geq0
\end{gathered}
\end{equation}

\point{Collateralized lending \& borrowing}
We consider a collateralized lending platform $\mathbb{L}$, which provides the \CollateralizedBorrow\ endpoint that requires the user to collateralize an asset $\mathsf{X}$ with a collateral factor $\mathsf{cf}$ (s.t. $0 < \mathsf{cf} < 1$) and borrows another asset $\mathsf{Y}$ at an exchange rate $\mathsf{er}$. The collateral factor determines the upper limit that a user can borrow. For example, if the collateral factor is $0.75$, a user is allowed to borrow up to 75\% of the value of the collateral. The exchange rate is for example determined by an outsourced price oracle. $z_{\mathsf{Y}}$ denotes the maximum amount of $\mathsf{Y}$ available for borrowing.

\states We hence consider the following state variables and ignore the balance changes of $\mathbb{L}$ for simplicity.
\begin{itemize}
    \item Balance of asset $\mathsf{X}$ held by $\mathbb{A}$: $\mathcal{B}(\mathbb{A};\mathsf{X};\mathsf{S})$
    \item Balance of asset $\mathsf{Y}$ held by $\mathbb{A}$: $\mathcal{B}(\mathbb{A};\mathsf{Y};\mathsf{S})$
\end{itemize}

\transitions The parameter $c_{\mathsf{X}}$ represents the amount of asset $\mathsf{X}$ that $\mathbb{A}$ aims to collateralize. Although $\mathbb{A}$ is allowed to borrow less than his collateral would allow for, we assume that $\mathbb{A}$ makes use the entirety of his collateral. Equation~\ref{eq:CollateralizedBorrow} shows the transition functions of \CollateralizedBorrow.

\begin{equation}\label{eq:CollateralizedBorrow}
\begin{gathered}
    \mathcal{B}(\mathbb{A};\mathsf{X};\mathsf{S'}) = \mathcal{B}(\mathbb{A};\mathsf{X};\mathsf{S}) - c_{\mathsf{X}}\\
    \mathcal{B}(\mathbb{A};\mathsf{Y};\mathsf{S'}) = \mathcal{B}(\mathbb{A};\mathsf{Y};\mathsf{S}) + b_{\mathsf{Y}}\\
    \text{where} \quad b_{\mathsf{Y}} = \frac{c_\mathsf{X}\times \mathsf{cf}}{\mathsf{er}}\\
    \text{s.t.} \quad \mathcal{B}(\mathbb{A};\mathsf{X};\mathsf{S'}) - c_{\mathsf{X}} \geq 0; z_{\mathsf{Y}} - b_{\mathsf{Y}} \geq 0
\end{gathered}
\end{equation}
$\mathbb{A}$ can retrieve its collateral by repaying the borrowed asset through the endpoint \CollateralizedRepay. We show the transition functions in Equation~\ref{eq:CollateralizedRepay} and for simplicity ignore the loan interest fee.
\begin{equation}\label{eq:CollateralizedRepay}
\begin{gathered}
    \mathcal{B}(\mathbb{A};\mathsf{X};\mathsf{S'}) = \mathcal{B}(\mathbb{A};\mathsf{X};\mathsf{S}) + c_{\mathsf{X}}\\
    \mathcal{B}(\mathbb{A};\mathsf{Y};\mathsf{S'}) = \mathcal{B}(\mathbb{A};\mathsf{Y};\mathsf{S}) - b_{\mathsf{Y}}\\
    \text{s.t.} \quad \mathcal{B}(\mathbb{A};\mathsf{Y};\mathsf{S}) - b_{\mathsf{Y}} \geq 0
\end{gathered}
\end{equation}

\point{Margin trading} 
A margin trading platform $\mathbb{T}$ allows the adversary $\mathbb{A}$ to short/long an asset $\mathsf{Y}$ by collateralizing asset $\mathsf{X}$ at a leverage $\ell$, where $\ell \geq 1$.

We focus on the \MarginShort\ endpoint which is relevant to the discussed DeFi attack in this work. We assume $\mathbb{A}$ shorts $\mathsf{Y}$ with respect to $\mathsf{X}$ on $\mathbb{F}$. 
The parameter $d_{\mathsf{X}}$ denotes the amount of $\mathsf{X}$ that $\mathsf{A}$ collateralizes upfront to open the margin. $w_\mathsf{X}$ represents the amount of $\mathsf{X}$ held by $\mathbb{F}$ that is available for the short margin. $\mathbb{A}$ is required to over-collateralize at a rate of $\mathsf{ocr}$ in a margin trade. In our model, when a short margin (short $\mathsf{Y}$ with respect to $\mathsf{X}$) is opened, $\mathbb{F}$ performs a trade on external $\mathsf{X}/\mathsf{Y}$ markets (e.g., Uniswap) to convert the leveraged $\mathsf{X}$ to $\mathsf{Y}$. The traded $\mathsf{Y}$ is locked until the margin is closed or liquidated.
\states In a short margin trading, we consider the following state variables:
\begin{itemize}
    \item Balance of $\mathsf{X}$ held by $\mathbb{A}$: $\mathcal{B}(\mathbb{A};\mathsf{X};\mathsf{S})$
    \item The locked amount of $\mathsf{Y}$: $\mathcal{L}(\mathbb{A};\mathsf{Y};\mathsf{S})$
\end{itemize}
\transitions We assume $\mathbb{F}$ transacts from an external market at a price of $\mathsf{emp}$. The transition functions and constraints are specified in Equation~\ref{eq:shortmargintrading}.
\begin{equation}\label{eq:shortmargintrading}
    \begin{gathered}
        \mathcal{B}(\mathbb{A};\mathsf{X};\mathsf{S'}) = \mathcal{B}(\mathbb{A};\mathsf{X};\mathsf{S}) - c_{\mathsf{X}}\\
        \mathcal{L}(\mathbb{A};\mathsf{Y};\mathsf{S}') = \mathcal{L}(\mathbb{A};\mathsf{Y};\mathsf{S}) + l_\mathsf{Y}\\
        \text{where}\quad l_{\mathsf{Y}}=\frac{d_{\mathsf{X}}\times \ell}{\mathsf{ocr}\times\mathsf{emp}}\\
        \text{s.t.}\quad\mathcal{B}(\mathbb{A};\mathsf{X};\mathsf{S}) - c_{\mathsf{X}} \geq 0; w_{\mathsf{X}} + d_{\mathsf{X}} - \frac{d_{\mathsf{X}}\times \ell}{\mathsf{ocr}} \geq 0
    \end{gathered}
\end{equation}

\section{Optimizing the Pump Attack and Arbitrage}\label{app:optimizingpumpandarbitrageattack}



In the following, we detail the procedure of deriving the pump attack and arbitrage optimization problem. Figure~\ref{fig:parametrization} summarizes the on-chain state when the attack was executed (i.e., $\mathsf{S}_0$). $\mathsf{X}$ and $\mathsf{Y}$ denote ETH and WBTC respectively. For simplicity, we ignore the trading fees in the constant product AMM (i.e., $\mathsf{f} = 0$ for $\mathbb{M}$). The endpoints executed in the pump attack and arbitrage are listed in the execution order as follows.
\begin{enumerate}[noitemsep]
    \item \Loan\ (dYdX)
    \item \CollateralizedBorrow\ (Compound)
    \item \MarginShort (bZx) \& \SwapXY\ (Uniswap)
    \item \SwapYX\ (Uniswap) 
    \item \Repay\ (dYdX)
    \item \SellXY\ \& \CollateralizedRepay (Compound)
\end{enumerate}
In the pump attack and arbitrage vector, we intend to tune the following two parameters, \textit{(i)} $p_1$: the amount of $\mathsf{X}$ collateralized to borrow $\mathsf{Y}$ in the endpoint 2) and \textit{(ii)} $p_2$: the amount of $\mathsf{X}$ collateralized to short $\mathsf{Y}$ in the endpoint 3). Following the procedure of Section~\ref{sec:parametrizedoptimization}, we proceed with detailing the construction of the constraint system.
\point{0)} We assume the initial balance of $\mathsf{X}$ owned by $\mathbb{A}$ is $\mathsf{B}_0$ (cf.\ Equation~\ref{eq:initial-balance}), and we refer the reader to Figure~\ref{fig:parametrization} for the remaining initial state values.
\begin{equation}\label{eq:initial-balance}
    \mathcal{B}(\mathbb{A};\mathsf{X};\mathsf{S}_0) = \mathsf{B}_0
\end{equation}

\point{1) \Loan}
$\mathbb{A}$ gets a flash loan of $\mathsf{X}$ amounts $p_1+p_2$ in total
\begin{equation*}
    \mathcal{B}(\mathbb{A};\mathsf{X};\mathsf{S}_1) = \mathsf{B}_0 + p_1+p_2
\end{equation*}
with the constraints
\begin{equation*}
\begin{gathered}
    p_1 \geq 0, p_2 \geq 0, v_{\mathsf{X}} - p_1 - p_2 \geq 0\\
\end{gathered}
\end{equation*}

\point{2) \CollateralizedBorrow}
$\mathbb{A}$ collateralizes $p_1$ amount of $\mathsf{X}$ to borrow $\mathsf{Y}$ from the lending platform $\mathbb{L}$
\begin{equation*}
\begin{gathered}
\mathcal{B}(\mathbb{A};\mathsf{X};\mathsf{S}_2) = \mathcal{B}(\mathbb{A};\mathsf{X};\mathsf{S}_1) - p_1 = \mathsf{B}_0 + p_2\\
\mathcal{B}(\mathbb{A};\mathsf{Y};\mathsf{S}_2) = \frac{p_1\times \mathsf{cf}}{\mathsf{er}}
\end{gathered}
\end{equation*}
\begin{equation*}
\mbox{with the constraint}~z_\mathsf{Y} - \frac{p_1\times \mathsf{cf}}{\mathsf{er}} \geq 0
\end{equation*}

\point{3) \MarginShort\ \& \SwapXY}
$\mathbb{A}$ opens a short margin with $p_2$ amount of $\mathsf{X}$ at a leverage of $\ell$ on the margin trading platform $\mathbb{T}$; $\mathbb{T}$ swaps the leveraged $\mathsf{X}$ for $\mathsf{Y}$ at the constant product AMM $\mathbb{M}$
\begin{equation*}
\begin{gathered}
\mathcal{B}(\mathbb{A};\mathsf{X};\mathsf{S}_3) = \mathcal{B}(\mathbb{A};\mathsf{X};\mathsf{S}_2) - p_2 = \mathsf{B}_0\\
u_\mathsf{X}(\mathsf{S}_3) = u_\mathsf{X}(\mathsf{S}_0) + \frac{p_2 \times \ell}{\mathsf{ocr}}\\
u_\mathsf{Y}(\mathsf{S}_3) = \frac{u_\mathsf{X}(\mathsf{S}_0)\times u_\mathsf{Y}(\mathsf{S}_0)}{u_\mathsf{X}(\mathsf{S}_3)}\\
\mathcal{L}(\mathbb{A};\mathsf{Y};\mathsf{S}_3) = u_\mathsf{Y}(\mathsf{S}_0) - u_\mathsf{Y}(\mathsf{S}_3)
\end{gathered}
\end{equation*}
\begin{equation*}
    \mbox{with the constraint}~w_\mathsf{X} + p_2 - \frac{p_2 \times \ell}{\mathsf{ocr}} \geq 0
\end{equation*}

\point{4) \SwapYX}
$\mathbb{A}$ dumps all the borrowed $\mathsf{Y}$ at $\mathbb{M}$
\begin{equation*}
\begin{gathered}
\mathcal{B}(\mathbb{A};\mathsf{Y};\mathsf{S}_4) = 0\\
u_\mathsf{Y}(\mathsf{S}_4) = u_\mathsf{Y}(\mathsf{S}_3) + \mathcal{B}(\mathbb{A};\mathsf{Y};\mathsf{S}_2)\\
u_\mathsf{X}(\mathsf{S}_4) = \frac{u_\mathsf{X}(\mathsf{S}_3)\times u_\mathsf{Y}(\mathsf{S}_3)}{u_\mathsf{Y}(\mathsf{S}_4)}\\
\mathcal{B}(\mathbb{A};\mathsf{X};\mathsf{S}_4) = \mathsf{B}_0 + u_\mathsf{X}(\mathsf{S}_3) - u_\mathsf{X}(\mathsf{S}_4)
\end{gathered}
\end{equation*}
\point{5) \Repay} 
$\mathbb{A}$ repays the flash loan
\begin{equation*}
\mathcal{B}(\mathbb{A};\mathsf{X};\mathsf{S}_5) = \mathcal{B}(\mathbb{A};\mathsf{X};\mathsf{S}_4) - p_1 - p_2
\end{equation*}
\begin{equation*}
\mbox{with the constraint}~\mathcal{B}(\mathbb{A};\mathsf{X};\mathsf{S}_4) - p_1 - p_2 \geq 0
\end{equation*}

\point{6) \SellXY\ \& \CollateralizedRepay}
$\mathbb{A}$ buys $Y$ from the market with the market price $\mathsf{p}_m$ and retrieves the collateral from $\mathbb{L}$
\begin{equation*}
\mathcal{B}(\mathbb{A};\mathsf{X};\mathsf{S}_6) =  \mathcal{B}(\mathbb{A};\mathsf{X};\mathsf{S}_5) + p_1 -  \mathcal{B}(\mathbb{A};\mathsf{Y};\mathsf{S}_2)\times\mathsf{p}_m
\end{equation*}

The objective function is the adversarial ETH revenue (cf.\ Equation~\ref{eq:pumparbitrageobject}).
\begin{equation}\label{eq:pumparbitrageobject}
\begin{aligned}
\mathcal{O}(\mathsf{S}_0; p_1; p_2) =\ &\mathcal{B}(\mathbb{A}; \mathsf{X}; \mathsf{S}_6) - \mathsf{B}_0\\
=\ &{u_\mathsf{X}(\mathsf{S}_0) + \frac{p_2 \times \ell}{\mathsf{ocr}}} - {u_\mathsf{X}(\mathsf{S}_4)} - p_2 \\
&- \frac{p_1\times \mathsf{cf} \times \mathsf{p}_m}{\mathsf{er}}
\end{aligned}
\end{equation}





\section{Optimizing the Oracle Manipulation Attack}\label{app:optimizingtheoraclemanipulationattack}
In the oracle manipulation attack, $\mathsf{X}$ denotes ETH and $\mathsf{Y}$ denotes sUSD. Again, we ignore the trading fees in the constant product AMM (i.e., $\mathsf{f} = 0$ for $\mathbb{M}$). The initial state variables are presented in Figure~\ref{fig:oraclemanipulationstate}. We assume that $\mathbb{A}$ owns zero balance of~$\mathsf{X}$ or~$\mathsf{Y}$. We list the endpoints involved in the oracle manipulation attack vector as follows.
\begin{enumerate}[noitemsep]
    \item \Loan (bZx)
    \item \SwapXY (Uniswap)
    \item \ConvertXY (Kyber reserve)
    \item \SellXY (Synthetix)
    \item \CollateralizedBorrow (bZx)
    \item \Repay (bZx)
\end{enumerate}



\noindent
We construct the constrained optimization problem as follows.
\point{1) \Loan}
$\mathbb{A}$ gets a flash loan of $\mathsf{X}$ amounts $p_1 + p_2 + p_3$
\begin{equation*}
    \mathcal{B}(\mathbb{A};\mathsf{X};\mathsf{S}_1) = p_1 + p_2 + p_3
\end{equation*}
with the constraints
\begin{equation*}
\begin{gathered}
    p_1 \geq 0, p_2 \geq 0, p_3 \geq 0, v_{\mathsf{X}} - p_1 - p_2 - p_3\geq 0
\end{gathered}
\end{equation*}

\point{2) \SwapXY}
$\mathbb{A}$ swaps $p_1$ amount of $\mathsf{X}$ for $\mathsf{Y}$ from the constant product AMM $\mathbb{M}$
\begin{equation*}
\begin{gathered}
\mathcal{B}(\mathbb{A};\mathsf{X};\mathsf{S}_2) = \mathcal{B}(\mathbb{A};\mathsf{X};\mathsf{S}_1) - p_1 = p_2 + p_3\\
u_\mathsf{X}(\mathsf{S}_2) = u_\mathsf{X}(\mathsf{S}_0) + p_1\\
u_\mathsf{Y}(\mathsf{S}_2) = \frac{u_\mathsf{X}(\mathsf{S}_0)\times u_\mathsf{Y}(\mathsf{S}_0)}{u_\mathsf{X}(\mathsf{S}_2)}\\
\mathcal{B}(\mathbb{A};\mathsf{Y};\mathsf{S}_2) = u_\mathsf{Y}(\mathsf{S}_0) - u_\mathsf{Y}(\mathsf{S}_2)
\end{gathered}
\end{equation*}

\point{3) \ConvertXY} $\mathbb{A}$ converts $p_2$ amount of $\mathsf{X}$ to $\mathsf{Y}$ from the automated price reserve $\mathbb{R}$
\begin{equation*}
\begin{gathered}
\mathcal{B}(\mathbb{A};\mathsf{X};\mathsf{S}_3)=\mathcal{B}(\mathbb{A};\mathsf{X};\mathsf{S}_2) - p_2 = p_1\\
k_\mathsf{X}(\mathsf{S}_3) = k_\mathsf{X}(\mathsf{S}_0) + p_2\\
\mathsf{P}_\mathsf{Y}(\mathbb{R};\mathsf{S}_3)=\mathsf{minP} \times e^{\mathsf{lr}\times k_\mathsf{X}(\mathsf{S}_3)}\\
\mathcal{B}(\mathbb{A};\mathsf{Y};\mathsf{S}_3)=\mathcal{B}(\mathbb{A};\mathsf{Y};\mathsf{S}_2) + \frac{e^{-\mathsf{lr}\times p_2} - 1}{\mathsf{lr}\times \mathsf{P}_\mathsf{Y}(\mathbb{R};\mathsf{S}_0)}\\
\text{s.t.} \quad \mathsf{maxP} - \mathsf{P}_\mathsf{Y}(\mathbb{R};\mathsf{S}_3)\geq 0
\end{gathered}
\end{equation*}

\point{4) \SellXY}
$\mathbb{A}$ sells $p_3$ amount of $\mathsf{X}$ for $\mathsf{Y}$ at the price of~$\mathsf{p}_m$
\begin{equation*}
\begin{gathered}
\mathcal{B}(\mathbb{A};\mathsf{X};\mathsf{S}_4) = \mathcal{B}(\mathbb{A};\mathsf{X};\mathsf{S}_3) - p_3 = 0\\
\mathcal{B}(\mathbb{A};\mathsf{Y};\mathsf{S}_4) = \mathcal{B}(\mathbb{A};\mathsf{Y};\mathsf{S}_3) + \frac{p_3}{\mathsf{p}_m}
\end{gathered}
\end{equation*}

\begin{equation*}
\mbox{with the constraint}~\mathsf{maxY} - \frac{p_3}{\mathsf{p}_m} \geq 0
\end{equation*}

\point{5) \CollateralizedBorrow}
$\mathbb{A}$ collateralizes all owned $\mathsf{Y}$ to borrow $\mathsf{X}$ according to the price given by the constant product AMM $\mathbb{M}$ (i.e., the exchange rate $\mathsf{er} = \frac{1}{\mathsf{P}_\mathsf{Y}(\mathbb{M};\mathsf{S}_2)}$)
\begin{equation*}
\begin{gathered}
\mathcal{B}(\mathbb{A};\mathsf{Y};\mathsf{S}_5) = 0\\
\mathcal{B}(\mathbb{A};\mathsf{X};\mathsf{S}_5) = \mathcal{B}(\mathbb{A};\mathsf{Y};\mathsf{S}_4)\times \mathsf{cf} \times \mathsf{P}_\mathsf{Y}(\mathbb{M};\mathsf{S}_2)
\end{gathered}
\end{equation*}
with the constraint
\begin{equation*}
    z_\mathsf{Y} - \mathcal{B}(\mathbb{A};\mathsf{Y};\mathsf{S}_4)\times \mathsf{cf} \times \mathsf{P}_\mathsf{Y}(\mathbb{M};\mathsf{S}_2) \geq 0
\end{equation*}

\point{6) \Repay}
$\mathbb{A}$ repays the flash loan
\begin{equation*}
    \mathcal{B}(\mathbb{A};\mathsf{X};\mathsf{S}_6)=\mathcal{B}(\mathbb{A};\mathsf{X};\mathsf{S}_5) - p_1 -p_2 -p_3
\end{equation*}
\begin{equation*}
    \mbox{with the constraint}~\mathcal{B}(\mathbb{A};\mathsf{X};\mathsf{S}_5) - p_1 -p_2 -p_3 \geq 0
\end{equation*}
The objective function is the remaining balance of $\mathsf{X}$ after repaying the flash loan (cf.\ Equation~\ref{eq:oracleobject}).
\begin{equation}\label{eq:oracleobject}
\begin{aligned}
\mathcal{O}(\mathsf{S}_0;p_1;p_2;p_3)=\ &\mathcal{B}(\mathbb{A};\mathsf{X};\mathsf{S}_6)\\
=\ & \mathcal{B}(\mathbb{A};\mathsf{X};\mathsf{S}_5) - p_1 -p_2 -p_3\\
=\ & \mathcal{B}(\mathbb{A};\mathsf{Y};\mathsf{S}_4)\times \mathsf{cf} \times \mathsf{P}_\mathsf{Y}(\mathbb{M};\mathsf{S}_2) \\
&- p_1 -p_2 -p_3
\end{aligned}
\end{equation}

\section{Extended Discussion}\label{app:extended-discussion}
In the following, we extend our discussion in Section~\ref{sec:discussion}.
\point{Responsible disclosure}
It is somewhat unclear how to perform responsible disclosure within DeFi, given that the underlying vulnerability and victim are not always perfectly clear and that there is a lack of security standards to apply. We plan to reach out to Aave, Kyber, and Uniswap to disclose the contents of this paper.

\point{Does extra capital help}
The main attraction of flash loans stems from them not requiring collateral that needs to be raised. One can, however, wonder whether extra capital would make the attacks we focus on more potent and the ROI greater. Based on our results,  extra collateral for the two attacks of Section~\ref{sec:postmortem} would not increase the ROI, as the liquidity constraints of the intermediate protocols do not allow for a higher impact.

\point{Potential defenses} 
Here we discuss several potential defenses. However, we would be the first to admit that these are not foolproof and come with potential downsides that would significantly hamper normal interactions.
\begin{itemize}
\item
Should DEX accept trades coming from flash loans?
\item
Should DEX accept coins from an address if the previous block did not show those funds in the address? 
\item
Would introducing a delay make sense, e.g., in governance voting, or price oracles?
\item
When designing a DeFi protocol, a single transaction should be limited in its abilities: a DEX should not allow a single transaction triggering a slippage beyond~100\%.
\end{itemize}


\point{Looking into the future}
In the future, we anticipate DeFi protocols eventually starting to comply with a higher standard of security testing, both within the protocol itself, as well as part of integration testing into the DeFi ecosystem. We believe that eventually, this may lead to some form of DeFi standards where it comes to financial security, similar to what is imposed on banks and other financial institutions in traditional centralized (government-controlled) finance. 
We anticipate that either whole-system penetration testing or an analytical approach to modeling the space of possibilities like in this paper are two ways to improve future DeFi protocols. 

\point{Generality of the optimization framework} We show in Section~\ref{sec:evaluation} that our optimization framework performs efficiently on a given attack vector. To discover new attacks on a blockchain state with the framework, we may need to iterate over all the combinations of DeFi actions. The search space thus explodes as the number of DeFi actions increases. Our optimization framework requires to model every DeFi action manually. This, however, makes the framework less handy for users who are unfamiliar with the mathematical formulas of the DeFi actions. To make the framework more accurate, we can build gas consumption and block gas limit into the models, which requires to comprehend every DeFi action explicitly. We leave the automation of modeling for future work.

\end{document}